\documentclass[aps,prl,letterpaper,amsmath,amssymb,superscriptaddress,nofootinbib,nolongbibliography,twocolumn]{revtex4-2}
\usepackage{graphicx}
\usepackage[usenames]{color}
\usepackage[colorlinks=true,linkcolor=cyan,citecolor=cyan,anchorcolor=cyan,urlcolor=cyan,bookmarks=true]{hyperref}

\usepackage{mathrsfs}
\usepackage{braket}

\usepackage{tabularx}
\usepackage{multirow}
\usepackage{orcidlink}
\usepackage{xcolor}

\begin{document}

\title{Discrete time crystal and perfect many-body tunneling in a periodically driven Heisenberg spin chain}
\author{Xiaotong Chen}
\affiliation{Tsung-Dao Lee Institute,
Shanghai Jiao Tong University, Shanghai, 201210, China}
\author{Jianda Wu}
\altaffiliation{wujd@sjtu.edu.cn}
\affiliation{Tsung-Dao Lee Institute,
Shanghai Jiao Tong University, Shanghai, 201210, China}
\affiliation{School of Physics \& Astronomy, Shanghai Jiao Tong University, Shanghai, 200240, China}
\affiliation{Shanghai Branch, Hefei National Laboratory, Shanghai 201315, China}

\begin{abstract}
We investigate the non-equilibrium dynamics of a Heisenberg spin-1/2 chain driven by a periodic magnetic field. 
Based on its instantaneous integrability and inherent symmetry, 
we analytically study the magnetization and many-body tunneling (MBT). 
Both of them exhibit periodicity distinct from the driving period. The magnetization is shown
to be independent of the initial state and robust against perturbations,
signaling the formation of discrete time crystal (DTC) order. 
The DTC phase is found to be continuously tunable through magnetic field.
The system exhibits perfect MBT, manifested as exactly vanishing Loschmidt echo (LE) thus 
divergent LE rate function 
at half period of the DTC. 
Remarkably, the perfect MBT is independent of the system size, and
can be traced to an effective gap closure induced by quantum geometric effects. 
Furthermore, the Loschmidt echo spectra entropy shows logarithmic-dependence on system size,
consistent with non-thermal nature of the DTC phase.
We propose a protocol using ultracold atoms for experimental realization of the DTC and MBT.
\end{abstract}
\maketitle

\textit{Introduction}---
Spontaneous symmetry breaking plays a key role in understanding phase transitions \cite{higgs_mechanism,SC_ssb,mass_anderson,bec_ssb,kt_ssb,subir}. A paradigmatic example 
is crystal formation—a phase transition from liquid to solid—in which continuous spatial translation symmetry is spontaneously broken.
Making an analogy to crystalline formation, Wilczek proposed ``time crystals" which  spontaneously 
break continuous time translation symmetry (TTS) \cite{frank,frank2}. Despite of
its difficulty to realize in equilibrium system \cite{masaki,no-go,bruno}, 
this kind of time crystal can still survive in peculiar systems \cite{ctc}. 
In contrast to the time crystal generated from continuous time translation symmetry breaking (TTSB), 
intensive studies on the periodically driven (Floquet) system
have revealed the existence of the discrete TTSB \cite{floquet, floquet1, floquet2, floquet3,dtc2,dtc3,dtc4,dtc5}:
Robust against the initial state choice and perturbations,
 a discrete time crystal (DTC) phase forms
with a time period distinct from the driving period of the system. 
Besides the the crystal order, a quasi-crystal, 
which is ordered without periodicity is also observed in a metallic solid \cite{realqc}. 
Drawing an analogy to quasi-crystal, a discrete time quasi-crystal (DTQC) has also been proposed, 
which are characterized by the presence of incommensurate spectra peaks of observables \cite{tqc1,tqc2,tqc3,tqc4}.

Under persistent driven, the system may continuously absorb energy and generally thermalize \cite{thermalize}. The requirement that some observables exhibit periodic behavior in the DTC phase imposes constraints on the viable systems. 
An example is the many body localization (MBL) system with emergent local conserved quantities \cite{floquet,mbl,mbl1,mbl2,mbl3},
where the DTCs stabilized by MBL have been proposed \cite{floquet, dtc_ls,nyyao_mbl_dtc_th}.
Another approach exploits prethermalization: 
an exponentially long prethermal time scale enables the DTC phase to be observed within this extended time window before eventual thermalization
\cite{prethermal, prethermalization2, prethermal4,prethermal5, prethermal6,prethermal7,prethermal8}.
These DTCs have been demonstrated in a variety of experimental platforms \cite{nyyao_mbl_dtc_exp, nyyao_mbl_dtc_exp2,processor_dtc, prethermal_exp1,prethermal_exp2}. 
In addition to these kinds of DTC, a quantum many body scar enabled DTC has been reported in kicked PXP model \cite{dtc_lukin}. 
However, this DTC phenomenon is not robust against initial state choice and only survives from a N{\'e}el-like initial state \cite{qmbs}. 
\par
Integrable systems, known as counterexample of eigenstate thermalization hypothesis, 
can also potentially host DTC phase \cite{review_qmbs}. However, 
general periodical driving typically breaks integrability.
In this letter, we study a Heisenberg spin$\text{-}1/2$ chain periodically driven by rotating magnetic field,
which keeps integrability instantaneously. 
We find a robust, initial-state-independent DTC phase with tunable period in the system. 
The robustness is manifested by the insensitivity of the response spectrum to small displacement field (DF) perturbation.
However, when the strength of the DF is comparable to that of rotating field, the DTC order is killed.
Instead, a possible DTQC phase is established. 
We further study the many-body tunneling (MBT) through 
the Loschmidt echo (LE) and its rate function which can
reveal global and local properties of the tunneling process, respectively.  
Interestingly, when the magnetic field parameters satisfy a simple constraint,
the system, independent of the size, remarkably exhibits vanishing LE thus divergent LE rate function 
at half period of the DTC phase indicating perfect MBT.   
The perfect MBT is attributed to an effective gap closure
due to quantum geometric effect.
Further investigation on the LE spectra entropy reveals its logarithmic 
dependence of the system size, consistent 
with the non-thermalized DTC phase. 
We further propose an experimental implementation of these non-equilibrium physics. 

\textit{Model.}---We start from an antiferromagnetic Heisenberg spin-1/2 
chain subjected to a periodic magnetic field
$\vec{h} (t)=h_r\mathrm{cos}(\omega t)
\hat{x}-h_r\mathrm{sin}(\omega t)\hat{y} -h\hat{z}$, 
\begin{equation}
    H(t)=H_{\text{xxx}}+\sum_{i=1}^{N} \vec{h}(t)\cdot\vec{S}_{i},
    \label{eq:H}
\end{equation}
where $H_{\text{xxx}} = J\sum_{i=1}^{N}\vec{S}_{i}\cdot\vec{S}_{i+1}$ with $J > 0$
and the spin vector $\vec{S}_{i}$ at site $i$. In addition, the periodic boundary
condition is imposed $\vec{S}_{N+1} = \vec{S}_{1}$.
The Hamiltonian has discrete TTS: $H(t)=H(t+T)$, $T=2\pi/\omega$.
Applying rotating wave transformation (RWT) $U_D(t)=\exp (-i\omega t\sum_{i=1}^{N} S_{i}^{z})$ \cite{wangxiaoprb},
we obtain  
\begin{equation}
  U_D(t)H(t)U_D(t)^{\dagger}=H(0),\;  \ket{\varphi_{m}(t)}=U_D(t)^{\dagger}\ket{\varphi_{m}(0)},
    \label{eq:basis}
\end{equation}
where $\ket{\varphi_{m}(t)}$ is the $m^{th}$ instantaneous eigenstate of $H(t)$ with instantaneous eigenenergy $E_m (t)$. Eq.~\eqref{eq:basis} 
implies the time independency of the spectrum of $H(t)$, namely $E_m(t)=E_m(0)$. 
With the help of $U_D(t)$, we can 
transform the Schrödinger equation
$i\partial_t \ket{\psi(t)}=H(t)\ket{\psi(t)}$ into
$i\partial_t \left(U_D(t)\ket{\psi(t)}\right)=H_{\text{eff}}\left(U_D(t)\ket{\psi(t)}\right)$
with
\begin{equation}
  H_{\text{eff}}=H(0)+\omega\sum_{j=1}^{N}S_{i}^{z}.
    \label{eq:heff}
\end{equation} 
Consequently, given a general initial state $\ket{\psi(0)}$, we have
$\ket{\psi(t)}=\mathcal{U}(t) \ket{\psi(0)}$ with $\mathcal{U}(t)\equiv U_D(t)^{\dagger}e^{-iH_{\text{eff}}t}$. 

\textit{DTC phase.}---
\begin{figure}
    \centering
    \includegraphics[width=0.98\linewidth]{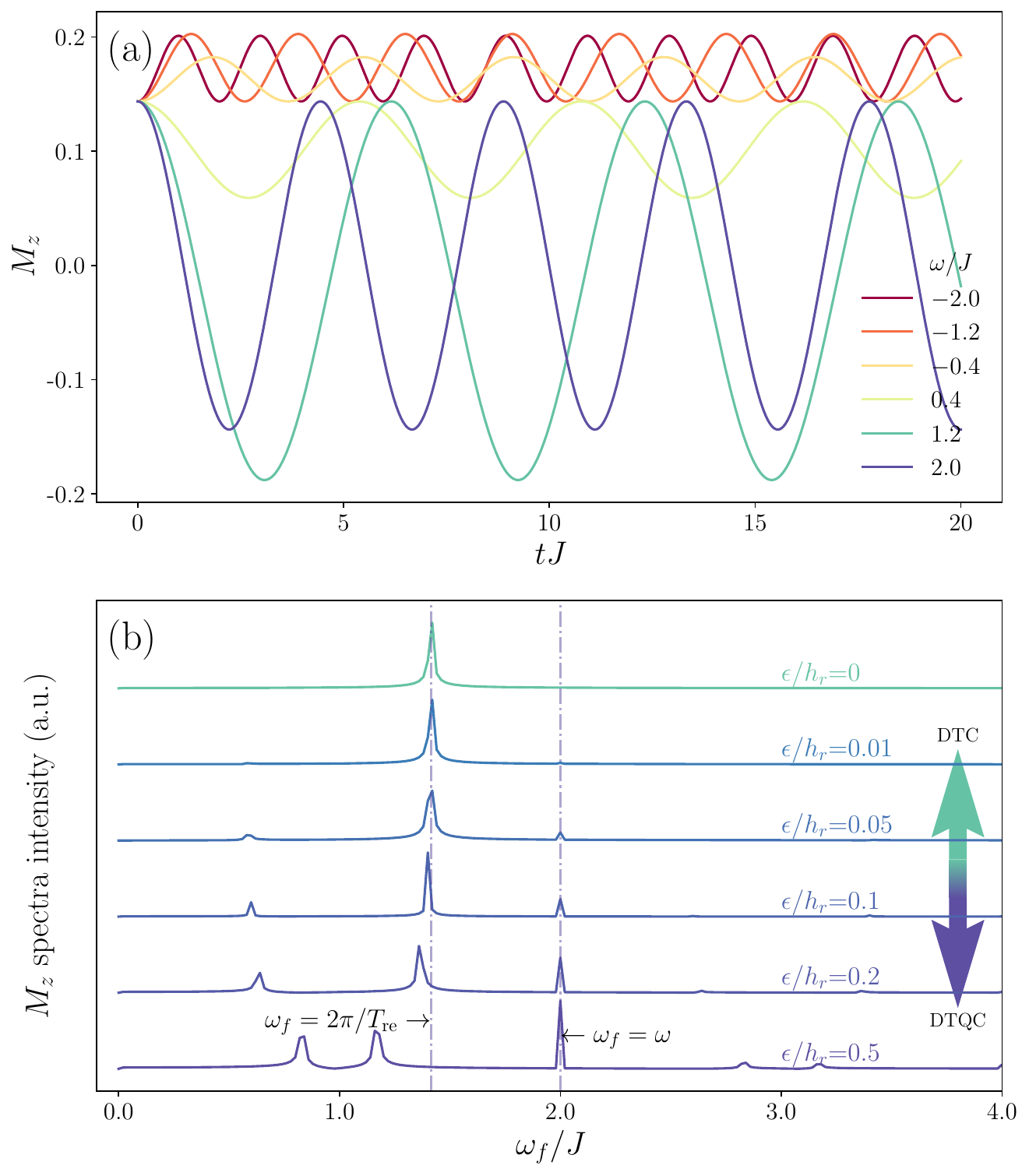}
    \caption{(a) $M_z$ $vs.$ $tJ$ with system size $N=64$ and magnetic field $h_r/J=h/J=1$. 
    $\omega/J$ ranges from $-$2 to 2 with stepwise 0.8. In each case, the $M_z$ exhibit periodic behavior. 
    (b) $M_z$ spectra intensity obtained through exact diagonalization with $h_r/J=h/J=1$, $\omega/J=2$, $N=12$ and different displacements $\epsilon$'s. 
    The initial state is $\ket{111110111110}$ with spin down $\ket{0}$ and spin up $\ket{1}$.
    Static magnetization background has been subtracted for all displacements.}
    \label{fig:fourier}
\end{figure}
The observables in this system exhibit periodic behavior with a period distinct from $T$. 
For example, we fix the ground state of $H(0)$ as an initial state determined by 
solving Bethe ansatz equations \cite{SM,Bethe}. Then the magnetization along the $z$ direction, $M_z$, 
is shown to host time period given by 
\begin{equation}
T_{\text{re}}=2\pi/\sqrt{h_r^2+(\omega-h)^2}
    \label{eq:Tre}
\end{equation} 
which is different from the driving period $T$ in general [Fig.~\ref{fig:fourier} (a)]. 
The period distinction indicates discrete TTSB thus the existence of the DTC phase.
To confirm the discrete TTSB, we need to further 
verify that the period 
distinction is independent of the initial state choice.  
For a general initial state $\ket{\varPhi}$, 
\begin{equation}
M_z=\frac{1}{N}\sum_{j}\bra{\varPhi}\mathcal{U}(t)^{\dagger}S_{j}^{z}\mathcal{U}(t)\ket{\varPhi}.
\end{equation}
Taking advantage of the SU(2) invariant of the $H_{\text{xxx}}$, we have $\mathcal{U}(t)^{\dagger}S_{j}^{z}\mathcal{U}(t)=\vec{f}(t)\cdot\vec{S}_{j}$, 
where $\vec{f}(t)$ is a periodic function with the same period as $T_{\text{re}}$ [Eq.~\ref{eq:Tre}] \cite{SM}. 
Consequently, $M_z = \vec{f}(t)\cdot \bra{\varPhi} \frac{1}{N}\sum_j\vec{S}_{j}\ket{\varPhi}$. 
The result confirms the existence of discrete TTSB.
Eq.~\eqref{eq:Tre} implies that the DTC period can be 
tuned freely by varying the magnetic field parameters.
The tunability stems from the continuous time 
dependence of our system. Since the RWT $U_D(t)$ commutes with $S_{j}^{z}$, 
then $M_z=\frac{1}{N}\sum_j\langle\chi(t)|S_{j}^{z}|\chi(t)\rangle$
with $|\chi(t)\rangle = e^{-iH_{\text{eff}}t}|\varPhi\rangle$,
which gives an effective Floquet unitary operator $U_F^{\text{eff}} \equiv
e^{-iH_{\text{eff}}T} = U_1U_2$
with $U_1=e^{-iT\sum_j[h_rS_{j}^{x}+(\omega-h)S_{j}^{z}]}$ and 
$U_2=e^{-iH_{\text{xxx}}T}$. Here
$U_1$ can be interpreted as an effective kick operator. 
Unlike kicked TTSB systems where the kick is explicitly 
designed in the time-dependent Hamiltonian, here the effective kick $U_1$ emerges 
from the applied magnetic field, leading to tunable periodicity of $M_z$ thus tunable
DTC period.  
A similar tunable DTC phase has also been proposed in one-dimensional periodically driven Hubbard models \cite{dtc5}.  

\par

An important feature of the DTC phase is its robustness against TTS-preserving perturbations.
Because a perturbation on the $z$-component of the field is irrelevant,
we consider a static displacement perturbation in the rotating plane,
\begin{equation}
\vec{h}(t)=\left[h_r\cos(\omega t)+\epsilon \right]\hat{x}-\left[h_r\sin(\omega t)-\epsilon\right]\hat{y}-h \hat{z}.
\end{equation}
The robustness of the DTC phase is manifested in the spectra intensity of $M_z$ [Fig.~\ref{fig:fourier} (b)]. 
When $|\epsilon/h_r| \ll1$ the dominant peak locates around $2\pi/T_{\text{re}}$.
However, when $\epsilon$ is comparable to $h_r$, 
the dominant peak shifts from $2\pi/T_{\text{re}}$ to the driving frequency $\omega$,
accompanied with additional satellite peaks. 
The ratio of the corresponding frequencies of these peaks 
to the driving frequency appear to be irrational.
Above process suggests a transition from DTC to DTQC tuned by the DF,
a new type of dynamical quantum phase transition.

\textit{Many-body tunneling.}---
\begin{figure}
    \centering
    \includegraphics[width=0.98\linewidth]{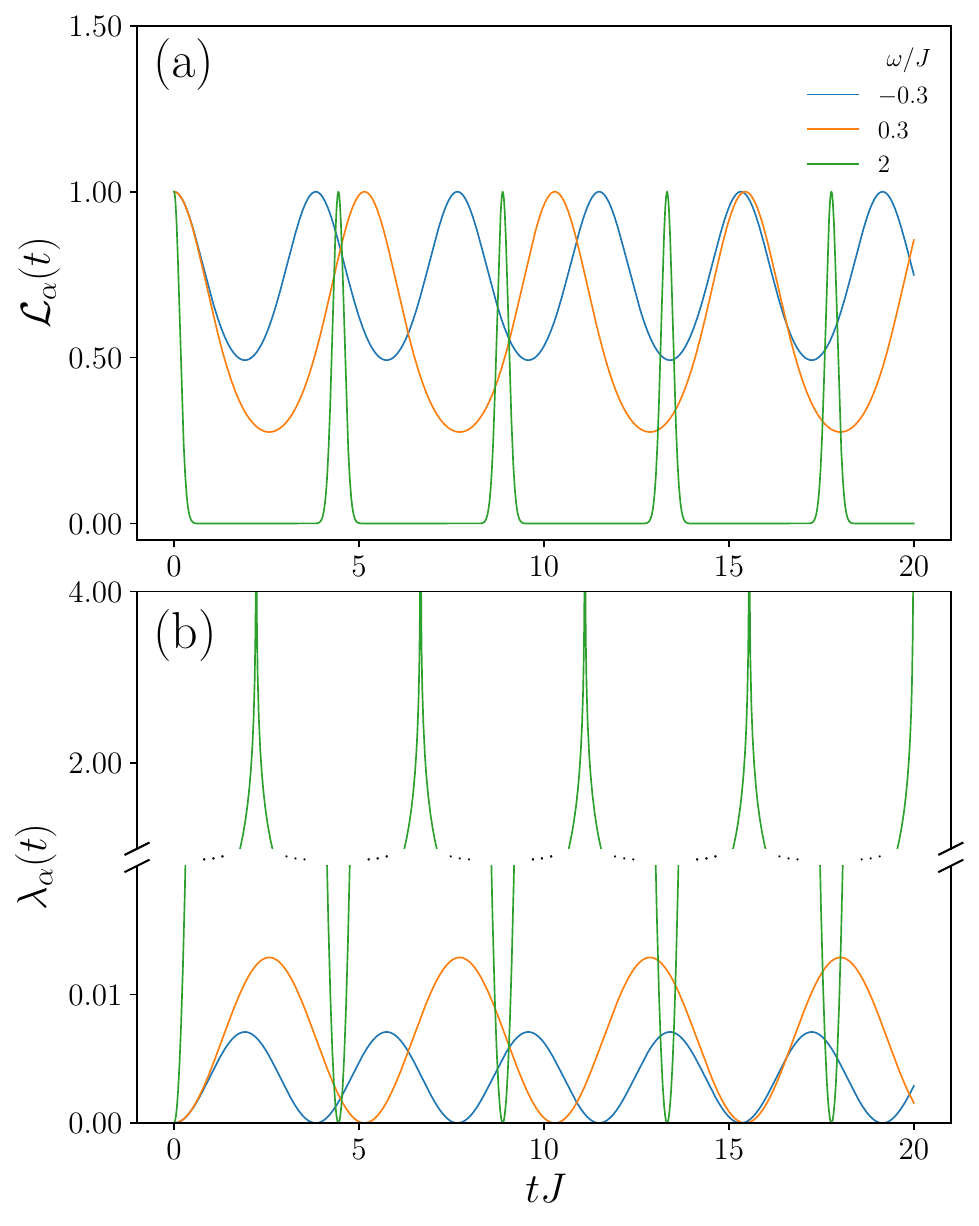}
    \caption{(a) LE {\it{vs.}} $tJ$ with system
     size $N=100$, magnetic field $h/J=h_r/J=1$, and different $\omega$'s.
    And $\omega/J=2$ 
     corresponds to $\alpha=\pi/2$. (b) LE rate function {\it{vs.}} $tJ$. 
     For $\omega/J=2$, $\lambda_{\alpha}(t)$ diverges at $tJ=\sqrt{2}(2n+1)\pi/2\ ,n\in\mathbb{N}$. }
    \label{fig:le}
\end{figure}
To investigate the MBT, we initialize the system in the ground state of $H(0)$, $\ket{GS(0)}$, then
introduce the Loschmidt echo (LE)
\begin{equation}
    \mathcal{L}(t)=\left|\braket{GS(t)|\psi(t)}\right|^2 = \left|\bra{GS(0)} e^{-iH_{\text{eff}}t}\ket{GS(0)}\right|^2
    \label{eq:le}
\end{equation}
to probe the MBT. Perfect MBT is characterized by the zeros of $\mathcal{L}(t)$.  
The eigenstates of $H_{\text{eff}}$ can be generated 
by applying a unitary transformation $U_\alpha=\text{exp}\left\{-i\alpha\sum_{j}S_{j}^{y}\right\}$ to the eigenstates of $H(0)$,
where $\tan\alpha=-h_r\omega/(h_r^2+h^2-h\omega)$. Furthermore, the eigenstates of $H(0)$ 
can be organized into SU(2) irreducible representations (IRs) 
where $\ket{GS(0)}$ carries the highest weight $\mathsf{j}$. 
By inserting complete basis of $H_{\text{eff}}$ in Eq.~\eqref{eq:le}, we have \cite{SM}
\begin{align}
\mathcal{L}_\alpha(t)&=\sum_{\mathsf{n=-j}}^{\mathsf{j}}\sum_{\mathsf{m=-j}}^{\mathsf{j}}\mathcal{O}_{\alpha,\mathsf{n}}\mathcal{O}_{\alpha,\mathsf{m}}\text{cos}(\tilde{h}(\mathsf{n-m})t)\\
&=\frac{\left(\tau^4+2\tau^2\cos\tilde{h}t+1\right)^{2\mathsf{j}}}{(1+\tau^2)^{4\mathsf{j}}},
    \label{eq:loschmidt_echo}
\end{align}
where $\tilde{h}\equiv\sqrt{h_r^2+(\omega-h)^2}$, $\tau=-\tan(\alpha/2)$ and $\mathcal{O}_{\alpha,\mathsf{n}}=\tau^{2(\mathsf{j-n})}C_{2\mathsf{j}}^{\mathsf{n+j}}/(1+\tau^2)^{2\mathsf{j}}$. 

Eq.~\eqref{eq:loschmidt_echo} gives the period of $\mathcal{L}_\alpha(t)$, which is identical to $T_{\text{re}}$ [Eq.~\ref{eq:Tre}]. 
Interestingly, when $\alpha = \pi/2$, Eq.~\eqref{eq:loschmidt_echo} can be further reduced to $\mathcal{L}_{\pi/2}(t)=\cos^{4\mathsf{j}}(\tilde{h}t/2)$, 
which directly implies perfect MBT at [Fig.~\ref{fig:le} (a)]
\begin{equation}
    t_l=(2l+1)\pi/\tilde{h} \equiv (2l+1)T_{\text{re}}/2,\;\; l \in \mathbb{Z}.
    \label{eq:tl}
\end{equation}
The special angle $\alpha=\pi/2$ can be fulfilled as long as
\begin{equation}
            h^2-h\omega+h_r^2=0
    \label{eq:square}
\end{equation}
holds.
The perfect MBT can be attributed to the quantum geometric effect. We introduce the 
instantaneous effective energy gap $\Delta^{\text{eff}}_{nm}$ between instantaneous 
eigenstate $\ket{\varphi_n(t)}$ and $\ket{\varphi_m(t)}$ \cite{jianda_qgp,geodesic_curvature,jianda_scipost},
\begin{equation}
    \Delta^{\text{eff}}_{nm}=E_n-E_m - Q_{nm} ,
    \label{eq:gap}
\end{equation}
where
$Q_{nm} = \mathcal{B}_{nn} - \mathcal{B}_{mm} - \frac{d}{dt}\text{arg}\ \mathcal{B}_{nm}$ is referred to as
quantum geometric potential which is invariant under local U(1) transformation $\ket{\varphi_n(t)} \to e^{i f_n(t) } \ket{\varphi_n(t)}$ and 
$\mathcal{B}_{nm}$ is the Berry connection matrix entry, $\mathcal{B}_{nm} = i\bra{\varphi_n(t)} \frac{d}{dt}\ket{\varphi_m(t)}$.
The perfect MBT appears when the effective energy gap closes, implying the two instantaneous states effectively degenerate although $E_n - E_m \neq 0$. 
For the model we study 
$\Delta^{\text{eff}}_{nm}=\bra{\varphi_n(0)}H_{\text{xxx}}\ket{\varphi_n(0)}-\bra{\varphi_m(0)}H_{\text{xxx}}\ket{\varphi_m(0)}$ when Eq.~\eqref{eq:square} 
holds. The SU(2) invariant of $H_{\text{xxx}}$ 
ensures existence of states within the same IR to make
$\Delta^{\text{eff}}_{nm}$ vanished. 

When magnetization density ($\mathsf{j}/N$) is fixed in the thermodynamic limit ($N \to \infty$), 
Eq.~\eqref{eq:loschmidt_echo}
indicates that the system only revives to the instantaneous ground state stroboscopically with period $T_{\text{re}}$, 
while tunneling into orthogonal states between revivals. This suggests, globally, perfect many-body tunneling (MBT) 
occurs almost everywhere for any nonzero $\alpha$, which is similar to the mechanism of orthogonality catastrophe \cite{anderson}. However, in local scenarios, the local-averaged tunneling rate characterized 
by the LE rate function $\lambda_\alpha (t) \equiv -\frac{1}{N}\ln\mathcal{L_\alpha}(t)$, 
quantifies the local deviation between $\ket{\psi(t)}$ and $\ket{GS(t)}$ \cite{LE}. 
Except at $t_l$, $\lambda_\alpha (t)$ remains finite, indicating that the state of 
a local spin does not fully tunnel to its orthogonal state on average. Conversely, 
the divergence of $\lambda_\alpha(t)$ at $t_l$ implies that, on average, 
the state of every local spin tunnels perfectly to its orthogonal one, 
implying perfect MBT even locally.

The physical picture of the perfect MBT is straightforward.
When $\alpha=\pi/2$, the direction of magnetic field of $H(0)$ is perpendicular 
to that of $H_{\text{eff}}$. The role of $H_{\text{eff}}$ is to drive $\ket{GS(0)}$ by a
magnetic field whose direction is perpendicular to the direction of magnetization of $\ket{GS(0)}$. 
After a certain time scale, namely $t_l$, the spins at every site are flipped, leading to perfect MBT. 
After completely tunneling to orthogonal space, the magnetic field drives the system back 
at $(l+1)T_{\text{re}}$, 
each spin is coherently driven back. Such physical picture accounts for the absence of thermalization.

\par
\begin{figure}
    \centering
    \includegraphics[width=0.98\columnwidth]{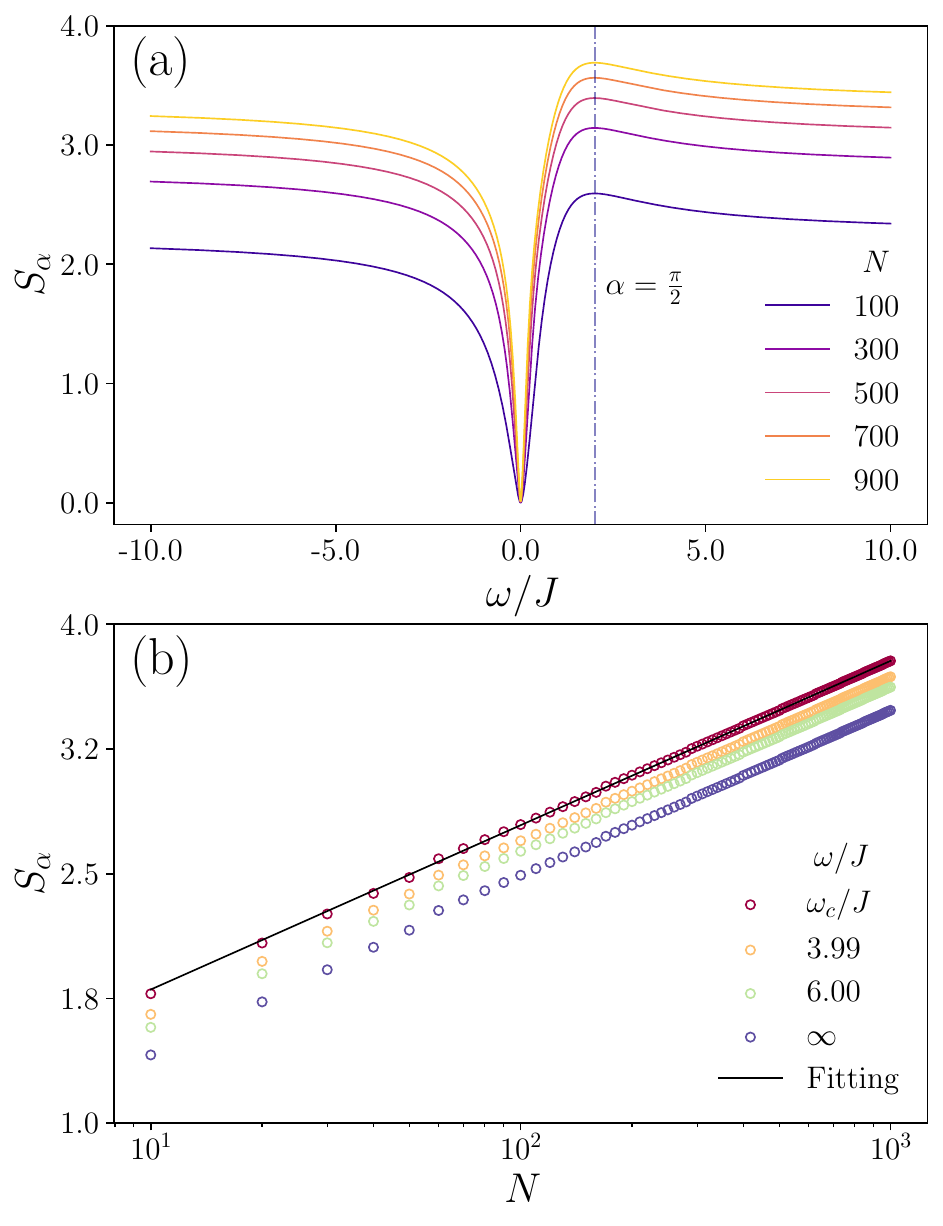}
    \caption{
    (a) SE $vs.$ $\omega$ for system size $N$ ranging from 100 to 900. (b) SE $vs.$ system size $N$,
    $S_{\alpha=\pi/2}\propto \kappa \ln N$, the fitting result shows that $\kappa=1.000(2)$.}

    \label{fig:se}
\end{figure}
We further study the spectrum entropy (SE) of LE \cite{spectrum_entropy}, 
$S_{\alpha} \equiv-\sum_{\mathsf{n m}} p_{\mathsf{\alpha n m}} \ln p_{\mathsf{\alpha n m}}$ with 
$p_{\mathsf{\alpha n m}}=\mathcal{O}_{\alpha,\mathsf{n}}\mathcal{O}_{\alpha,\mathsf{m}}$ given by 
$2\pi\sum_{\mathsf{n m}} p_{\mathsf{\alpha n m}}\delta(\omega-\omega_{\mathsf{n m}})=\int_{-\infty}^{\infty}e^{-i \omega t}\mathcal{L}_{\alpha}(t)dt$.
Then the LE spectrum entropy follows
\begin{equation}
    S_{\alpha} =-2\sum_{\mathsf{n=-j}}^{\mathsf{j}}\mathcal{O}_{\alpha, \mathsf{n}}\ln \mathcal{O}_{\alpha, \mathsf{n}}.
    \label{eq:SE}
\end{equation}
$S_\alpha$ above is equivalent to twice the von Neumann entropy of a mixed state 
with density matrix given by $\rho=\sum_{\mathsf{n}}\mathcal{O}_{\alpha,\mathsf{n}}\ket{\mathsf{j,n}}\bra{\mathsf{j,n}}$. 
This shows that the SE can shed light in understanding the time evolution process for given $\omega$ and $N$. 
The SE is determined by the distribution of $\mathcal{O}_{\alpha, \mathsf{n}}$ w.r.t $\mathsf{n}$. 
$S_\alpha$ reaches its maximum value when $\alpha=\pi/2$ [Fig.~\ref{fig:se}(a)] under which the perfect MBT occurs.
 A logarithmic behavior $S_{\alpha = \pi/2}\propto\kappa\ln N$ emerges [Fig.~\ref{fig:se}(b)]. 
The $\kappa\ln N$ behavior originates from the symmetry constraint: the number 
of nonzero matrix elements of $\rho$ is determined by the magnetization of $\ket{GS(0)}$, which increases linearly as the system size increases. 
Such logarithmic behavior breaks down in the static limit $\omega\rightarrow0$, 
in which $H_{\text{eff}}$ can be regarded as  $H(0)$ perturbed by a tilted field, 
leading to non-vanishing $\mathcal{O}_{\alpha,\mathsf{n}}$ only when $\mathsf{n}\approx\mathsf{j}$. 

\par
\textit{Ultracold atoms platform implementation.}---
Recently, a preparation of a ground state of the antiferromagnetic 
Heisenberg chain has been achieved in ultracold atom platforms 
described by a two-component Bose Hubbard model \cite{cold_atom}. 
The progress makes it possible for an experimental implementation for exploring the DTC phase 
as well as MBT using the cold atoms platform. 
The static field in $z$ direction and the 
in-plane rotating field
can be realized via species-dependent potential and the microwave,
respectively. Together, they give rise to the desired time-dependent coupling
$\Omega\sum_{j}[\cos\phi(t)S_{j}^{x}-\sin\phi(t)S_{j}^{y}]-\delta\sum_j S_{j}^{z}$ 
with $\Omega$, $\phi(t)$ and $\delta$ being the Rabi frequency, phase of microwave, 
and potential difference, respectively. 
 
\par
For probing the MBT, initially the phase is fixed at $\phi(0)=0$, and 
the system is prepared in the ground state. 
Set $\phi(t)=\omega t$ with $\omega$ satisfying $\delta^2-\delta\omega+\Omega^2=0$. 
During the time evolution, the $M_z$ and 
the spin correlation can be measured by fluorescence imaging. 
$M_z$ is expected to oscillate with period $T_{\text{re}}=2\pi/\sqrt{\Omega^2+(\omega-\delta)^2}$, 
while the period of spin correlation will be half of that of magnetization, which is due to the $\mathbb{Z}_2$ symmetry \cite{SM}. 
The perfect MBT can be identified by verifying the characteristic 
periods $T_{\text{re}}$ and $T_{\text{re}}/2$ for $M_z$ and spin correlation, respectively. 
\par
Ground state preparation is unnecessary for DTC observation. 
And no constraints are imposed on $\omega$.
For any initial state with nonzero magnetization, after the time-dependent phase $\phi(t)=\omega t$ is turned on,  
the DTC phenomenon can be identified by measuring the periodic behavior of $M_z$. 
Meanwhile for the Néel-like initial state with zero magnetization, 
the measurement of staggered magnetization $\frac{1}{N}\braket{\sum_{j}(-1)^{j}S_{j}^{z}}$ serves as indicator of the DTC phenomenon. 
To investigate possible DTC-DTQC transition, 
an additional microwave with Rabi frequency $\Omega_{\text{DF}}$ and constant phase is introduced 
to realize the DF. 
By tuning $\Omega_{\text{DF}}$, such phenomenon can be observed. 

\textit{Discussion and Conclusion.}---
We demonstrated a DTC phase in a time-period driven Heisenberg chain. 
The robustness against DF perturbation has been verified. 
The DF perturbation we consider leads to small residual time-dependent term in the rotating frame. 
Whether this term can lead to a prethermalization 
time scale requires further investigation. 
Interestingly, large DF leads to dominant peak located at 
driving frequency accompanied with comparable satellite peaks, 
potentially implying a dynamical quantum phase transition from DTC to DTQC. 
It is worth to explore the mechanism of this transition and determine the transition point. 
\par
The logarithmic maximization of SE w.r.t system size when perfect MBT occurs 
reflects the complexity of the tunneling processes. 
Symmetry constraints restrict the dimension of the accessible subspace, 
thereby determining the system size dependence of SE. 
The imposition of SU(2) symmetry dictates that the maximal SE growth is logarithmic: $S_\alpha\sim\kappa\ln N$ with $\kappa\leq2$, where $\kappa$ varies between different cases.   
To illustrate the upper bound of $\kappa$, consider the infinite high temperature mixed state 
where $\mathcal{O}_{\alpha,\mathsf{n}} = 1/(2\mathsf{j}+1)$ for all $\mathsf{n} \in {-\mathsf{j}, \ldots, \mathsf{j}}$. 
This uniform distribution gives rise to $S_{\alpha} = 2\ln(2\mathsf{j}+1) \sim 2\ln N$.

To summarize, we identified a robust and
initial-state-independent DTC phase in a rotating-field-driven Heisenberg chain, 
whose response period is tunable. The tunability arises from the SU(2) invariant of the $H_{\text{xxx}}$
such that $[U_1,U_2]=0$ for any choice of $h$ and $h_r$. 
We also demonstrated the perfect MBT, which is attributed to the closure of the effective energy gap. 
The quantum geometric effect leads to an effective degeneracy between the states split by magnetic field, 
signaling an emergent SU(2) invariant with respect to the effective energy level. 
Our study shows that the quantum geometric effect plays an important role in
understanding quantum many body non-equilibrium dynamics. 
We expect our results to be tested in a suitable cold-atom experimental setup.

\textit{Acknowledgements.}---We want to thank S. Liu, J. Yang, X. Wang, Y. Gao, H. Sun, D. Yang, K. Liu
for helpful discussions. The work is
sponsored by National Natural Science Foundation of
China Nos. 12274288, 12450004, the Innovation Program for Quantum Science and Technology Grant No.
2021ZD0301900.

\bibliography{manuscript_time_driven_heisenberg}

\begin{thebibliography}{59}%
\makeatletter
\providecommand \@ifxundefined [1]{%
 \@ifx{#1\undefined}
}%
\providecommand \@ifnum [1]{%
 \ifnum #1\expandafter \@firstoftwo
 \else \expandafter \@secondoftwo
 \fi
}%
\providecommand \@ifx [1]{%
 \ifx #1\expandafter \@firstoftwo
 \else \expandafter \@secondoftwo
 \fi
}%
\providecommand \natexlab [1]{#1}%
\providecommand \enquote  [1]{``#1''}%
\providecommand \bibnamefont  [1]{#1}%
\providecommand \bibfnamefont [1]{#1}%
\providecommand \citenamefont [1]{#1}%
\providecommand \href@noop [0]{\@secondoftwo}%
\providecommand \href [0]{\begingroup \@sanitize@url \@href}%
\providecommand \@href[1]{\@@startlink{#1}\@@href}%
\providecommand \@@href[1]{\endgroup#1\@@endlink}%
\providecommand \@sanitize@url [0]{\catcode `\\12\catcode `\$12\catcode
  `\&12\catcode `\#12\catcode `\^12\catcode `\_12\catcode `\%12\relax}%
\providecommand \@@startlink[1]{}%
\providecommand \@@endlink[0]{}%
\providecommand \url  [0]{\begingroup\@sanitize@url \@url }%
\providecommand \@url [1]{\endgroup\@href {#1}{\urlprefix }}%
\providecommand \urlprefix  [0]{URL }%
\providecommand \Eprint [0]{\href }%
\providecommand \doibase [0]{https://doi.org/}%
\providecommand \selectlanguage [0]{\@gobble}%
\providecommand \bibinfo  [0]{\@secondoftwo}%
\providecommand \bibfield  [0]{\@secondoftwo}%
\providecommand \translation [1]{[#1]}%
\providecommand \BibitemOpen [0]{}%
\providecommand \bibitemStop [0]{}%
\providecommand \bibitemNoStop [0]{.\EOS\space}%
\providecommand \EOS [0]{\spacefactor3000\relax}%
\providecommand \BibitemShut  [1]{\csname bibitem#1\endcsname}%
\let\auto@bib@innerbib\@empty
\bibitem [{\citenamefont {Higgs}(1964)}]{higgs_mechanism}%
  \BibitemOpen
  \bibfield  {author} {\bibinfo {author} {\bibfnamefont {P.~W.}\ \bibnamefont
  {Higgs}},\ }\href {https://doi.org/10.1103/PhysRevLett.13.508} {\bibfield
  {journal} {\bibinfo  {journal} {Phys. Rev. Lett.}\ }\textbf {\bibinfo
  {volume} {13}},\ \bibinfo {pages} {508} (\bibinfo {year} {1964})}\BibitemShut
  {NoStop}%
\bibitem [{\citenamefont {Sigrist}\ and\ \citenamefont {Ueda}(1991)}]{SC_ssb}%
  \BibitemOpen
  \bibfield  {author} {\bibinfo {author} {\bibfnamefont {M.}~\bibnamefont
  {Sigrist}}\ and\ \bibinfo {author} {\bibfnamefont {K.}~\bibnamefont {Ueda}},\
  }\href {https://doi.org/10.1103/RevModPhys.63.239} {\bibfield  {journal}
  {\bibinfo  {journal} {Rev. Mod. Phys.}\ }\textbf {\bibinfo {volume} {63}},\
  \bibinfo {pages} {239} (\bibinfo {year} {1991})}\BibitemShut {NoStop}%
\bibitem [{\citenamefont {Anderson}(1963)}]{mass_anderson}%
  \BibitemOpen
  \bibfield  {author} {\bibinfo {author} {\bibfnamefont {P.~W.}\ \bibnamefont
  {Anderson}},\ }\href {https://doi.org/10.1103/PhysRev.130.439} {\bibfield
  {journal} {\bibinfo  {journal} {Phys. Rev.}\ }\textbf {\bibinfo {volume}
  {130}},\ \bibinfo {pages} {439} (\bibinfo {year} {1963})}\BibitemShut
  {NoStop}%
\bibitem [{\citenamefont {Lieb}\ \emph {et~al.}(2007)\citenamefont {Lieb},
  \citenamefont {Seiringer},\ and\ \citenamefont {Yngvason}}]{bec_ssb}%
  \BibitemOpen
  \bibfield  {author} {\bibinfo {author} {\bibfnamefont {E.~H.}\ \bibnamefont
  {Lieb}}, \bibinfo {author} {\bibfnamefont {R.}~\bibnamefont {Seiringer}},\
  and\ \bibinfo {author} {\bibfnamefont {J.}~\bibnamefont {Yngvason}},\ }\href
  {https://doi.org/https://doi.org/10.1016/S0034-4877(07)80074-7} {\bibfield
  {journal} {\bibinfo  {journal} {Reports on Mathematical Physics}\ }\textbf
  {\bibinfo {volume} {59}},\ \bibinfo {pages} {389} (\bibinfo {year}
  {2007})}\BibitemShut {NoStop}%
\bibitem [{\citenamefont {Kosterlitz}\ and\ \citenamefont
  {Thouless}(1973)}]{kt_ssb}%
  \BibitemOpen
  \bibfield  {author} {\bibinfo {author} {\bibfnamefont {J.~M.}\ \bibnamefont
  {Kosterlitz}}\ and\ \bibinfo {author} {\bibfnamefont {D.~J.}\ \bibnamefont
  {Thouless}},\ }\href {https://doi.org/10.1088/0022-3719/6/7/010} {\bibfield
  {journal} {\bibinfo  {journal} {Journal of Physics C: Solid State Physics}\
  }\textbf {\bibinfo {volume} {6}},\ \bibinfo {pages} {1181} (\bibinfo {year}
  {1973})}\BibitemShut {NoStop}%
\bibitem [{\citenamefont {Sachdev}(2011)}]{subir}%
  \BibitemOpen
  \bibfield  {author} {\bibinfo {author} {\bibfnamefont {S.}~\bibnamefont
  {Sachdev}},\ }\href@noop {} {\emph {\bibinfo {title} {Quantum Phase
  Transitions}}},\ \bibinfo {edition} {2nd}\ ed.\ (\bibinfo  {publisher}
  {Cambridge University Press},\ \bibinfo {year} {2011})\BibitemShut {NoStop}%
\bibitem [{\citenamefont {Wilczek}(2012)}]{frank}%
  \BibitemOpen
  \bibfield  {author} {\bibinfo {author} {\bibfnamefont {F.}~\bibnamefont
  {Wilczek}},\ }\href {https://doi.org/10.1103/PhysRevLett.109.160401}
  {\bibfield  {journal} {\bibinfo  {journal} {Phys. Rev. Lett.}\ }\textbf
  {\bibinfo {volume} {109}},\ \bibinfo {pages} {160401} (\bibinfo {year}
  {2012})}\BibitemShut {NoStop}%
\bibitem [{\citenamefont {Wilczek}(2013)}]{frank2}%
  \BibitemOpen
  \bibfield  {author} {\bibinfo {author} {\bibfnamefont {F.}~\bibnamefont
  {Wilczek}},\ }\href {https://doi.org/10.1103/PhysRevLett.111.250402}
  {\bibfield  {journal} {\bibinfo  {journal} {Phys. Rev. Lett.}\ }\textbf
  {\bibinfo {volume} {111}},\ \bibinfo {pages} {250402} (\bibinfo {year}
  {2013})}\BibitemShut {NoStop}%
\bibitem [{\citenamefont {Watanabe}\ and\ \citenamefont
  {Oshikawa}(2015)}]{masaki}%
  \BibitemOpen
  \bibfield  {author} {\bibinfo {author} {\bibfnamefont {H.}~\bibnamefont
  {Watanabe}}\ and\ \bibinfo {author} {\bibfnamefont {M.}~\bibnamefont
  {Oshikawa}},\ }\href {https://doi.org/10.1103/PhysRevLett.114.251603}
  {\bibfield  {journal} {\bibinfo  {journal} {Phys. Rev. Lett.}\ }\textbf
  {\bibinfo {volume} {114}},\ \bibinfo {pages} {251603} (\bibinfo {year}
  {2015})}\BibitemShut {NoStop}%
\bibitem [{\citenamefont {Bruno}(2013{\natexlab{a}})}]{no-go}%
  \BibitemOpen
  \bibfield  {author} {\bibinfo {author} {\bibfnamefont {P.}~\bibnamefont
  {Bruno}},\ }\href {https://doi.org/10.1103/PhysRevLett.111.070402} {\bibfield
   {journal} {\bibinfo  {journal} {Phys. Rev. Lett.}\ }\textbf {\bibinfo
  {volume} {111}},\ \bibinfo {pages} {070402} (\bibinfo {year}
  {2013}{\natexlab{a}})}\BibitemShut {NoStop}%
\bibitem [{\citenamefont {Bruno}(2013{\natexlab{b}})}]{bruno}%
  \BibitemOpen
  \bibfield  {author} {\bibinfo {author} {\bibfnamefont {P.}~\bibnamefont
  {Bruno}},\ }\href {https://doi.org/10.1103/PhysRevLett.110.118901} {\bibfield
   {journal} {\bibinfo  {journal} {Phys. Rev. Lett.}\ }\textbf {\bibinfo
  {volume} {110}},\ \bibinfo {pages} {118901} (\bibinfo {year}
  {2013}{\natexlab{b}})}\BibitemShut {NoStop}%
\bibitem [{\citenamefont {Kozin}\ and\ \citenamefont {Kyriienko}(2019)}]{ctc}%
  \BibitemOpen
  \bibfield  {author} {\bibinfo {author} {\bibfnamefont {V.~K.}\ \bibnamefont
  {Kozin}}\ and\ \bibinfo {author} {\bibfnamefont {O.}~\bibnamefont
  {Kyriienko}},\ }\href {https://doi.org/10.1103/PhysRevLett.123.210602}
  {\bibfield  {journal} {\bibinfo  {journal} {Phys. Rev. Lett.}\ }\textbf
  {\bibinfo {volume} {123}},\ \bibinfo {pages} {210602} (\bibinfo {year}
  {2019})}\BibitemShut {NoStop}%
\bibitem [{\citenamefont {Else}\ \emph {et~al.}(2016)\citenamefont {Else},
  \citenamefont {Bauer},\ and\ \citenamefont {Nayak}}]{floquet}%
  \BibitemOpen
  \bibfield  {author} {\bibinfo {author} {\bibfnamefont {D.~V.}\ \bibnamefont
  {Else}}, \bibinfo {author} {\bibfnamefont {B.}~\bibnamefont {Bauer}},\ and\
  \bibinfo {author} {\bibfnamefont {C.}~\bibnamefont {Nayak}},\ }\href
  {https://doi.org/10.1103/PhysRevLett.117.090402} {\bibfield  {journal}
  {\bibinfo  {journal} {Phys. Rev. Lett.}\ }\textbf {\bibinfo {volume} {117}},\
  \bibinfo {pages} {090402} (\bibinfo {year} {2016})}\BibitemShut {NoStop}%
\bibitem [{\citenamefont {von Keyserlingk}\ \emph {et~al.}(2016)\citenamefont
  {von Keyserlingk}, \citenamefont {Khemani},\ and\ \citenamefont
  {Sondhi}}]{floquet1}%
  \BibitemOpen
  \bibfield  {author} {\bibinfo {author} {\bibfnamefont {C.~W.}\ \bibnamefont
  {von Keyserlingk}}, \bibinfo {author} {\bibfnamefont {V.}~\bibnamefont
  {Khemani}},\ and\ \bibinfo {author} {\bibfnamefont {S.~L.}\ \bibnamefont
  {Sondhi}},\ }\href {https://doi.org/10.1103/PhysRevB.94.085112} {\bibfield
  {journal} {\bibinfo  {journal} {Phys. Rev. B}\ }\textbf {\bibinfo {volume}
  {94}},\ \bibinfo {pages} {085112} (\bibinfo {year} {2016})}\BibitemShut
  {NoStop}%
\bibitem [{\citenamefont {Khemani}\ \emph {et~al.}(2016)\citenamefont
  {Khemani}, \citenamefont {Lazarides}, \citenamefont {Moessner},\ and\
  \citenamefont {Sondhi}}]{floquet2}%
  \BibitemOpen
  \bibfield  {author} {\bibinfo {author} {\bibfnamefont {V.}~\bibnamefont
  {Khemani}}, \bibinfo {author} {\bibfnamefont {A.}~\bibnamefont {Lazarides}},
  \bibinfo {author} {\bibfnamefont {R.}~\bibnamefont {Moessner}},\ and\
  \bibinfo {author} {\bibfnamefont {S.~L.}\ \bibnamefont {Sondhi}},\ }\href
  {https://doi.org/10.1103/PhysRevLett.116.250401} {\bibfield  {journal}
  {\bibinfo  {journal} {Phys. Rev. Lett.}\ }\textbf {\bibinfo {volume} {116}},\
  \bibinfo {pages} {250401} (\bibinfo {year} {2016})}\BibitemShut {NoStop}%
\bibitem [{\citenamefont {von Keyserlingk}\ and\ \citenamefont
  {Sondhi}(2016)}]{floquet3}%
  \BibitemOpen
  \bibfield  {author} {\bibinfo {author} {\bibfnamefont {C.~W.}\ \bibnamefont
  {von Keyserlingk}}\ and\ \bibinfo {author} {\bibfnamefont {S.~L.}\
  \bibnamefont {Sondhi}},\ }\href {https://doi.org/10.1103/PhysRevB.93.245146}
  {\bibfield  {journal} {\bibinfo  {journal} {Phys. Rev. B}\ }\textbf {\bibinfo
  {volume} {93}},\ \bibinfo {pages} {245146} (\bibinfo {year}
  {2016})}\BibitemShut {NoStop}%
\bibitem [{\citenamefont {Russomanno}\ \emph {et~al.}(2017)\citenamefont
  {Russomanno}, \citenamefont {Iemini}, \citenamefont {Dalmonte},\ and\
  \citenamefont {Fazio}}]{dtc2}%
  \BibitemOpen
  \bibfield  {author} {\bibinfo {author} {\bibfnamefont {A.}~\bibnamefont
  {Russomanno}}, \bibinfo {author} {\bibfnamefont {F.}~\bibnamefont {Iemini}},
  \bibinfo {author} {\bibfnamefont {M.}~\bibnamefont {Dalmonte}},\ and\
  \bibinfo {author} {\bibfnamefont {R.}~\bibnamefont {Fazio}},\ }\href
  {https://doi.org/10.1103/PhysRevB.95.214307} {\bibfield  {journal} {\bibinfo
  {journal} {Phys. Rev. B}\ }\textbf {\bibinfo {volume} {95}},\ \bibinfo
  {pages} {214307} (\bibinfo {year} {2017})}\BibitemShut {NoStop}%
\bibitem [{\citenamefont {Gong}\ \emph {et~al.}(2018)\citenamefont {Gong},
  \citenamefont {Hamazaki},\ and\ \citenamefont {Ueda}}]{dtc3}%
  \BibitemOpen
  \bibfield  {author} {\bibinfo {author} {\bibfnamefont {Z.}~\bibnamefont
  {Gong}}, \bibinfo {author} {\bibfnamefont {R.}~\bibnamefont {Hamazaki}},\
  and\ \bibinfo {author} {\bibfnamefont {M.}~\bibnamefont {Ueda}},\ }\href
  {https://doi.org/10.1103/PhysRevLett.120.040404} {\bibfield  {journal}
  {\bibinfo  {journal} {Phys. Rev. Lett.}\ }\textbf {\bibinfo {volume} {120}},\
  \bibinfo {pages} {040404} (\bibinfo {year} {2018})}\BibitemShut {NoStop}%
\bibitem [{\citenamefont {Huang}\ \emph {et~al.}(2018)\citenamefont {Huang},
  \citenamefont {Wu},\ and\ \citenamefont {Liu}}]{dtc4}%
  \BibitemOpen
  \bibfield  {author} {\bibinfo {author} {\bibfnamefont {B.}~\bibnamefont
  {Huang}}, \bibinfo {author} {\bibfnamefont {Y.-H.}\ \bibnamefont {Wu}},\ and\
  \bibinfo {author} {\bibfnamefont {W.~V.}\ \bibnamefont {Liu}},\ }\href
  {https://doi.org/10.1103/PhysRevLett.120.110603} {\bibfield  {journal}
  {\bibinfo  {journal} {Phys. Rev. Lett.}\ }\textbf {\bibinfo {volume} {120}},\
  \bibinfo {pages} {110603} (\bibinfo {year} {2018})}\BibitemShut {NoStop}%
\bibitem [{\citenamefont {Chinzei}\ and\ \citenamefont {Ikeda}(2020)}]{dtc5}%
  \BibitemOpen
  \bibfield  {author} {\bibinfo {author} {\bibfnamefont {K.}~\bibnamefont
  {Chinzei}}\ and\ \bibinfo {author} {\bibfnamefont {T.~N.}\ \bibnamefont
  {Ikeda}},\ }\href {https://doi.org/10.1103/PhysRevLett.125.060601} {\bibfield
   {journal} {\bibinfo  {journal} {Phys. Rev. Lett.}\ }\textbf {\bibinfo
  {volume} {125}},\ \bibinfo {pages} {060601} (\bibinfo {year}
  {2020})}\BibitemShut {NoStop}%
\bibitem [{\citenamefont {Shechtman}\ \emph {et~al.}(1984)\citenamefont
  {Shechtman}, \citenamefont {Blech}, \citenamefont {Gratias},\ and\
  \citenamefont {Cahn}}]{realqc}%
  \BibitemOpen
  \bibfield  {author} {\bibinfo {author} {\bibfnamefont {D.}~\bibnamefont
  {Shechtman}}, \bibinfo {author} {\bibfnamefont {I.}~\bibnamefont {Blech}},
  \bibinfo {author} {\bibfnamefont {D.}~\bibnamefont {Gratias}},\ and\ \bibinfo
  {author} {\bibfnamefont {J.~W.}\ \bibnamefont {Cahn}},\ }\href
  {https://doi.org/10.1103/PhysRevLett.53.1951} {\bibfield  {journal} {\bibinfo
   {journal} {Phys. Rev. Lett.}\ }\textbf {\bibinfo {volume} {53}},\ \bibinfo
  {pages} {1951} (\bibinfo {year} {1984})}\BibitemShut {NoStop}%
\bibitem [{\citenamefont {Dumitrescu}\ \emph {et~al.}(2018)\citenamefont
  {Dumitrescu}, \citenamefont {Vasseur},\ and\ \citenamefont {Potter}}]{tqc1}%
  \BibitemOpen
  \bibfield  {author} {\bibinfo {author} {\bibfnamefont {P.~T.}\ \bibnamefont
  {Dumitrescu}}, \bibinfo {author} {\bibfnamefont {R.}~\bibnamefont
  {Vasseur}},\ and\ \bibinfo {author} {\bibfnamefont {A.~C.}\ \bibnamefont
  {Potter}},\ }\href {https://doi.org/10.1103/PhysRevLett.120.070602}
  {\bibfield  {journal} {\bibinfo  {journal} {Phys. Rev. Lett.}\ }\textbf
  {\bibinfo {volume} {120}},\ \bibinfo {pages} {070602} (\bibinfo {year}
  {2018})}\BibitemShut {NoStop}%
\bibitem [{\citenamefont {Giergiel}\ \emph {et~al.}(2018)\citenamefont
  {Giergiel}, \citenamefont {Miroszewski},\ and\ \citenamefont {Sacha}}]{tqc2}%
  \BibitemOpen
  \bibfield  {author} {\bibinfo {author} {\bibfnamefont {K.}~\bibnamefont
  {Giergiel}}, \bibinfo {author} {\bibfnamefont {A.}~\bibnamefont
  {Miroszewski}},\ and\ \bibinfo {author} {\bibfnamefont {K.}~\bibnamefont
  {Sacha}},\ }\href {https://doi.org/10.1103/PhysRevLett.120.140401} {\bibfield
   {journal} {\bibinfo  {journal} {Phys. Rev. Lett.}\ }\textbf {\bibinfo
  {volume} {120}},\ \bibinfo {pages} {140401} (\bibinfo {year}
  {2018})}\BibitemShut {NoStop}%
\bibitem [{\citenamefont {Hou}\ \emph {et~al.}(2018)\citenamefont {Hou},
  \citenamefont {Hu}, \citenamefont {Sun},\ and\ \citenamefont {Zhang}}]{tqc3}%
  \BibitemOpen
  \bibfield  {author} {\bibinfo {author} {\bibfnamefont {J.}~\bibnamefont
  {Hou}}, \bibinfo {author} {\bibfnamefont {H.}~\bibnamefont {Hu}}, \bibinfo
  {author} {\bibfnamefont {K.}~\bibnamefont {Sun}},\ and\ \bibinfo {author}
  {\bibfnamefont {C.}~\bibnamefont {Zhang}},\ }\href
  {https://doi.org/10.1103/PhysRevLett.120.060407} {\bibfield  {journal}
  {\bibinfo  {journal} {Phys. Rev. Lett.}\ }\textbf {\bibinfo {volume} {120}},\
  \bibinfo {pages} {060407} (\bibinfo {year} {2018})}\BibitemShut {NoStop}%
\bibitem [{\citenamefont {Else}\ \emph {et~al.}(2020)\citenamefont {Else},
  \citenamefont {Ho},\ and\ \citenamefont {Dumitrescu}}]{tqc4}%
  \BibitemOpen
  \bibfield  {author} {\bibinfo {author} {\bibfnamefont {D.~V.}\ \bibnamefont
  {Else}}, \bibinfo {author} {\bibfnamefont {W.~W.}\ \bibnamefont {Ho}},\ and\
  \bibinfo {author} {\bibfnamefont {P.~T.}\ \bibnamefont {Dumitrescu}},\ }\href
  {https://doi.org/10.1103/PhysRevX.10.021032} {\bibfield  {journal} {\bibinfo
  {journal} {Phys. Rev. X}\ }\textbf {\bibinfo {volume} {10}},\ \bibinfo
  {pages} {021032} (\bibinfo {year} {2020})}\BibitemShut {NoStop}%
\bibitem [{\citenamefont {D'Alessio}\ and\ \citenamefont
  {Rigol}(2014)}]{thermalize}%
  \BibitemOpen
  \bibfield  {author} {\bibinfo {author} {\bibfnamefont {L.}~\bibnamefont
  {D'Alessio}}\ and\ \bibinfo {author} {\bibfnamefont {M.}~\bibnamefont
  {Rigol}},\ }\href {https://doi.org/10.1103/PhysRevX.4.041048} {\bibfield
  {journal} {\bibinfo  {journal} {Phys. Rev. X}\ }\textbf {\bibinfo {volume}
  {4}},\ \bibinfo {pages} {041048} (\bibinfo {year} {2014})}\BibitemShut
  {NoStop}%
\bibitem [{\citenamefont {Serbyn}\ \emph {et~al.}(2013)\citenamefont {Serbyn},
  \citenamefont {Papi\ifmmode~\acute{c}\else \'{c}\fi{}},\ and\ \citenamefont
  {Abanin}}]{mbl}%
  \BibitemOpen
  \bibfield  {author} {\bibinfo {author} {\bibfnamefont {M.}~\bibnamefont
  {Serbyn}}, \bibinfo {author} {\bibfnamefont {Z.}~\bibnamefont
  {Papi\ifmmode~\acute{c}\else \'{c}\fi{}}},\ and\ \bibinfo {author}
  {\bibfnamefont {D.~A.}\ \bibnamefont {Abanin}},\ }\href
  {https://doi.org/10.1103/PhysRevLett.111.127201} {\bibfield  {journal}
  {\bibinfo  {journal} {Phys. Rev. Lett.}\ }\textbf {\bibinfo {volume} {111}},\
  \bibinfo {pages} {127201} (\bibinfo {year} {2013})}\BibitemShut {NoStop}%
\bibitem [{\citenamefont {Ponte}\ \emph
  {et~al.}(2015{\natexlab{a}})\citenamefont {Ponte}, \citenamefont {Chandran},
  \citenamefont {Papić},\ and\ \citenamefont {Abanin}}]{mbl1}%
  \BibitemOpen
  \bibfield  {author} {\bibinfo {author} {\bibfnamefont {P.}~\bibnamefont
  {Ponte}}, \bibinfo {author} {\bibfnamefont {A.}~\bibnamefont {Chandran}},
  \bibinfo {author} {\bibfnamefont {Z.}~\bibnamefont {Papić}},\ and\ \bibinfo
  {author} {\bibfnamefont {D.~A.}\ \bibnamefont {Abanin}},\ }\href
  {https://doi.org/https://doi.org/10.1016/j.aop.2014.11.008} {\bibfield
  {journal} {\bibinfo  {journal} {Annals of Physics}\ }\textbf {\bibinfo
  {volume} {353}},\ \bibinfo {pages} {196} (\bibinfo {year}
  {2015}{\natexlab{a}})}\BibitemShut {NoStop}%
\bibitem [{\citenamefont {Ponte}\ \emph
  {et~al.}(2015{\natexlab{b}})\citenamefont {Ponte}, \citenamefont
  {Papi\ifmmode~\acute{c}\else \'{c}\fi{}}, \citenamefont {Huveneers},\ and\
  \citenamefont {Abanin}}]{mbl2}%
  \BibitemOpen
  \bibfield  {author} {\bibinfo {author} {\bibfnamefont {P.}~\bibnamefont
  {Ponte}}, \bibinfo {author} {\bibfnamefont {Z.}~\bibnamefont
  {Papi\ifmmode~\acute{c}\else \'{c}\fi{}}}, \bibinfo {author} {\bibfnamefont
  {F.~m.~c.}\ \bibnamefont {Huveneers}},\ and\ \bibinfo {author} {\bibfnamefont
  {D.~A.}\ \bibnamefont {Abanin}},\ }\href
  {https://doi.org/10.1103/PhysRevLett.114.140401} {\bibfield  {journal}
  {\bibinfo  {journal} {Phys. Rev. Lett.}\ }\textbf {\bibinfo {volume} {114}},\
  \bibinfo {pages} {140401} (\bibinfo {year} {2015}{\natexlab{b}})}\BibitemShut
  {NoStop}%
\bibitem [{\citenamefont {Lazarides}\ \emph {et~al.}(2015)\citenamefont
  {Lazarides}, \citenamefont {Das},\ and\ \citenamefont {Moessner}}]{mbl3}%
  \BibitemOpen
  \bibfield  {author} {\bibinfo {author} {\bibfnamefont {A.}~\bibnamefont
  {Lazarides}}, \bibinfo {author} {\bibfnamefont {A.}~\bibnamefont {Das}},\
  and\ \bibinfo {author} {\bibfnamefont {R.}~\bibnamefont {Moessner}},\ }\href
  {https://doi.org/10.1103/PhysRevLett.115.030402} {\bibfield  {journal}
  {\bibinfo  {journal} {Phys. Rev. Lett.}\ }\textbf {\bibinfo {volume} {115}},\
  \bibinfo {pages} {030402} (\bibinfo {year} {2015})}\BibitemShut {NoStop}%
\bibitem [{\citenamefont {Liu}\ \emph {et~al.}(2023)\citenamefont {Liu},
  \citenamefont {Zhang}, \citenamefont {Hsieh}, \citenamefont {Zhang},\ and\
  \citenamefont {Yao}}]{dtc_ls}%
  \BibitemOpen
  \bibfield  {author} {\bibinfo {author} {\bibfnamefont {S.}~\bibnamefont
  {Liu}}, \bibinfo {author} {\bibfnamefont {S.-X.}\ \bibnamefont {Zhang}},
  \bibinfo {author} {\bibfnamefont {C.-Y.}\ \bibnamefont {Hsieh}}, \bibinfo
  {author} {\bibfnamefont {S.}~\bibnamefont {Zhang}},\ and\ \bibinfo {author}
  {\bibfnamefont {H.}~\bibnamefont {Yao}},\ }\href
  {https://doi.org/10.1103/PhysRevLett.130.120403} {\bibfield  {journal}
  {\bibinfo  {journal} {Phys. Rev. Lett.}\ }\textbf {\bibinfo {volume} {130}},\
  \bibinfo {pages} {120403} (\bibinfo {year} {2023})}\BibitemShut {NoStop}%
\bibitem [{\citenamefont {Yao}\ \emph {et~al.}(2017)\citenamefont {Yao},
  \citenamefont {Potter}, \citenamefont {Potirniche},\ and\ \citenamefont
  {Vishwanath}}]{nyyao_mbl_dtc_th}%
  \BibitemOpen
  \bibfield  {author} {\bibinfo {author} {\bibfnamefont {N.~Y.}\ \bibnamefont
  {Yao}}, \bibinfo {author} {\bibfnamefont {A.~C.}\ \bibnamefont {Potter}},
  \bibinfo {author} {\bibfnamefont {I.-D.}\ \bibnamefont {Potirniche}},\ and\
  \bibinfo {author} {\bibfnamefont {A.}~\bibnamefont {Vishwanath}},\ }\href
  {https://doi.org/10.1103/PhysRevLett.118.030401} {\bibfield  {journal}
  {\bibinfo  {journal} {Phys. Rev. Lett.}\ }\textbf {\bibinfo {volume} {118}},\
  \bibinfo {pages} {030401} (\bibinfo {year} {2017})}\BibitemShut {NoStop}%
\bibitem [{\citenamefont {Else}\ \emph {et~al.}(2017)\citenamefont {Else},
  \citenamefont {Bauer},\ and\ \citenamefont {Nayak}}]{prethermal}%
  \BibitemOpen
  \bibfield  {author} {\bibinfo {author} {\bibfnamefont {D.~V.}\ \bibnamefont
  {Else}}, \bibinfo {author} {\bibfnamefont {B.}~\bibnamefont {Bauer}},\ and\
  \bibinfo {author} {\bibfnamefont {C.}~\bibnamefont {Nayak}},\ }\href
  {https://doi.org/10.1103/PhysRevX.7.011026} {\bibfield  {journal} {\bibinfo
  {journal} {Phys. Rev. X}\ }\textbf {\bibinfo {volume} {7}},\ \bibinfo {pages}
  {011026} (\bibinfo {year} {2017})}\BibitemShut {NoStop}%
\bibitem [{\citenamefont {Luitz}\ \emph {et~al.}(2020)\citenamefont {Luitz},
  \citenamefont {Moessner}, \citenamefont {Sondhi},\ and\ \citenamefont
  {Khemani}}]{prethermalization2}%
  \BibitemOpen
  \bibfield  {author} {\bibinfo {author} {\bibfnamefont {D.~J.}\ \bibnamefont
  {Luitz}}, \bibinfo {author} {\bibfnamefont {R.}~\bibnamefont {Moessner}},
  \bibinfo {author} {\bibfnamefont {S.~L.}\ \bibnamefont {Sondhi}},\ and\
  \bibinfo {author} {\bibfnamefont {V.}~\bibnamefont {Khemani}},\ }\href
  {https://doi.org/10.1103/PhysRevX.10.021046} {\bibfield  {journal} {\bibinfo
  {journal} {Phys. Rev. X}\ }\textbf {\bibinfo {volume} {10}},\ \bibinfo
  {pages} {021046} (\bibinfo {year} {2020})}\BibitemShut {NoStop}%
\bibitem [{\citenamefont {Mori}\ \emph
  {et~al.}(2016{\natexlab{a}})\citenamefont {Mori}, \citenamefont {Kuwahara},\
  and\ \citenamefont {Saito}}]{prethermal4}%
  \BibitemOpen
  \bibfield  {author} {\bibinfo {author} {\bibfnamefont {T.}~\bibnamefont
  {Mori}}, \bibinfo {author} {\bibfnamefont {T.}~\bibnamefont {Kuwahara}},\
  and\ \bibinfo {author} {\bibfnamefont {K.}~\bibnamefont {Saito}},\ }\href
  {https://doi.org/10.1103/PhysRevLett.116.120401} {\bibfield  {journal}
  {\bibinfo  {journal} {Phys. Rev. Lett.}\ }\textbf {\bibinfo {volume} {116}},\
  \bibinfo {pages} {120401} (\bibinfo {year} {2016}{\natexlab{a}})}\BibitemShut
  {NoStop}%
\bibitem [{\citenamefont {Chandran}\ and\ \citenamefont
  {Sondhi}(2016)}]{prethermal5}%
  \BibitemOpen
  \bibfield  {author} {\bibinfo {author} {\bibfnamefont {A.}~\bibnamefont
  {Chandran}}\ and\ \bibinfo {author} {\bibfnamefont {S.~L.}\ \bibnamefont
  {Sondhi}},\ }\href {https://doi.org/10.1103/PhysRevB.93.174305} {\bibfield
  {journal} {\bibinfo  {journal} {Phys. Rev. B}\ }\textbf {\bibinfo {volume}
  {93}},\ \bibinfo {pages} {174305} (\bibinfo {year} {2016})}\BibitemShut
  {NoStop}%
\bibitem [{\citenamefont {Abanin}\ \emph
  {et~al.}(2017{\natexlab{a}})\citenamefont {Abanin}, \citenamefont {De~Roeck},
  \citenamefont {Ho},\ and\ \citenamefont {Huveneers}}]{prethermal6}%
  \BibitemOpen
  \bibfield  {author} {\bibinfo {author} {\bibfnamefont {D.~A.}\ \bibnamefont
  {Abanin}}, \bibinfo {author} {\bibfnamefont {W.}~\bibnamefont {De~Roeck}},
  \bibinfo {author} {\bibfnamefont {W.~W.}\ \bibnamefont {Ho}},\ and\ \bibinfo
  {author} {\bibfnamefont {F.~m.~c.}\ \bibnamefont {Huveneers}},\ }\href
  {https://doi.org/10.1103/PhysRevB.95.014112} {\bibfield  {journal} {\bibinfo
  {journal} {Phys. Rev. B}\ }\textbf {\bibinfo {volume} {95}},\ \bibinfo
  {pages} {014112} (\bibinfo {year} {2017}{\natexlab{a}})}\BibitemShut
  {NoStop}%
\bibitem [{\citenamefont {Abanin}\ \emph
  {et~al.}(2017{\natexlab{b}})\citenamefont {Abanin}, \citenamefont {De~Roeck},
  \citenamefont {Ho},\ and\ \citenamefont {Huveneers}}]{prethermal7}%
  \BibitemOpen
  \bibfield  {author} {\bibinfo {author} {\bibfnamefont {D.}~\bibnamefont
  {Abanin}}, \bibinfo {author} {\bibfnamefont {W.}~\bibnamefont {De~Roeck}},
  \bibinfo {author} {\bibfnamefont {W.~W.}\ \bibnamefont {Ho}},\ and\ \bibinfo
  {author} {\bibfnamefont {F.}~\bibnamefont {Huveneers}},\ }\href
  {https://doi.org/10.1007/s00220-017-2930-x} {\bibfield  {journal} {\bibinfo
  {journal} {Communications in Mathematical Physics}\ }\textbf {\bibinfo
  {volume} {354}},\ \bibinfo {pages} {809} (\bibinfo {year}
  {2017}{\natexlab{b}})}\BibitemShut {NoStop}%
\bibitem [{\citenamefont {Mori}\ \emph
  {et~al.}(2016{\natexlab{b}})\citenamefont {Mori}, \citenamefont {Kuwahara},\
  and\ \citenamefont {Saito}}]{prethermal8}%
  \BibitemOpen
  \bibfield  {author} {\bibinfo {author} {\bibfnamefont {T.}~\bibnamefont
  {Mori}}, \bibinfo {author} {\bibfnamefont {T.}~\bibnamefont {Kuwahara}},\
  and\ \bibinfo {author} {\bibfnamefont {K.}~\bibnamefont {Saito}},\ }\href
  {https://doi.org/10.1103/PhysRevLett.116.120401} {\bibfield  {journal}
  {\bibinfo  {journal} {Phys. Rev. Lett.}\ }\textbf {\bibinfo {volume} {116}},\
  \bibinfo {pages} {120401} (\bibinfo {year} {2016}{\natexlab{b}})}\BibitemShut
  {NoStop}%
\bibitem [{\citenamefont {Zhang}\ \emph {et~al.}(2017)\citenamefont {Zhang},
  \citenamefont {Hess}, \citenamefont {Kyprianidis}, \citenamefont {Becker},
  \citenamefont {Lee}, \citenamefont {Smith}, \citenamefont {Pagano},
  \citenamefont {Potirniche}, \citenamefont {Potter}, \citenamefont
  {Vishwanath}, \citenamefont {Yao},\ and\ \citenamefont
  {Monroe}}]{nyyao_mbl_dtc_exp}%
  \BibitemOpen
  \bibfield  {author} {\bibinfo {author} {\bibfnamefont {J.}~\bibnamefont
  {Zhang}}, \bibinfo {author} {\bibfnamefont {P.~W.}\ \bibnamefont {Hess}},
  \bibinfo {author} {\bibfnamefont {A.}~\bibnamefont {Kyprianidis}}, \bibinfo
  {author} {\bibfnamefont {P.}~\bibnamefont {Becker}}, \bibinfo {author}
  {\bibfnamefont {A.}~\bibnamefont {Lee}}, \bibinfo {author} {\bibfnamefont
  {J.}~\bibnamefont {Smith}}, \bibinfo {author} {\bibfnamefont
  {G.}~\bibnamefont {Pagano}}, \bibinfo {author} {\bibfnamefont {I.-D.}\
  \bibnamefont {Potirniche}}, \bibinfo {author} {\bibfnamefont {A.~C.}\
  \bibnamefont {Potter}}, \bibinfo {author} {\bibfnamefont {A.}~\bibnamefont
  {Vishwanath}}, \bibinfo {author} {\bibfnamefont {N.~Y.}\ \bibnamefont
  {Yao}},\ and\ \bibinfo {author} {\bibfnamefont {C.}~\bibnamefont {Monroe}},\
  }\href {https://doi.org/https://doi.org/10.1038/nature21413} {\bibfield
  {journal} {\bibinfo  {journal} {Nature}\ }\textbf {\bibinfo {volume} {543}},\
  \bibinfo {pages} {217} (\bibinfo {year} {2017})}\BibitemShut {NoStop}%
\bibitem [{\citenamefont {Choi}\ \emph {et~al.}(2017)\citenamefont {Choi},
  \citenamefont {Choi}, \citenamefont {Landig}, \citenamefont {Kucsko},
  \citenamefont {Zhou}, \citenamefont {Isoya}, \citenamefont {Jelezko},
  \citenamefont {Onoda}, \citenamefont {Sumiya}, \citenamefont {Khemani},
  \citenamefont {von Keyserlingk}, \citenamefont {Yao}, \citenamefont
  {Demler},\ and\ \citenamefont {Lukin}}]{nyyao_mbl_dtc_exp2}%
  \BibitemOpen
  \bibfield  {author} {\bibinfo {author} {\bibfnamefont {S.}~\bibnamefont
  {Choi}}, \bibinfo {author} {\bibfnamefont {J.}~\bibnamefont {Choi}}, \bibinfo
  {author} {\bibfnamefont {R.}~\bibnamefont {Landig}}, \bibinfo {author}
  {\bibfnamefont {G.}~\bibnamefont {Kucsko}}, \bibinfo {author} {\bibfnamefont
  {H.}~\bibnamefont {Zhou}}, \bibinfo {author} {\bibfnamefont {J.}~\bibnamefont
  {Isoya}}, \bibinfo {author} {\bibfnamefont {F.}~\bibnamefont {Jelezko}},
  \bibinfo {author} {\bibfnamefont {S.}~\bibnamefont {Onoda}}, \bibinfo
  {author} {\bibfnamefont {H.}~\bibnamefont {Sumiya}}, \bibinfo {author}
  {\bibfnamefont {V.}~\bibnamefont {Khemani}}, \bibinfo {author} {\bibfnamefont
  {C.}~\bibnamefont {von Keyserlingk}}, \bibinfo {author} {\bibfnamefont
  {N.~Y.}\ \bibnamefont {Yao}}, \bibinfo {author} {\bibfnamefont
  {E.}~\bibnamefont {Demler}},\ and\ \bibinfo {author} {\bibfnamefont {M.~D.}\
  \bibnamefont {Lukin}},\ }\href
  {https://doi.org/https://doi.org/10.1038/nature21426} {\bibfield  {journal}
  {\bibinfo  {journal} {Nature}\ }\textbf {\bibinfo {volume} {543}},\ \bibinfo
  {pages} {221} (\bibinfo {year} {2017})}\BibitemShut {NoStop}%
\bibitem [{\citenamefont {Mi}\ \emph {et~al.}(2022)\citenamefont {Mi} \emph
  {et~al.}}]{processor_dtc}%
  \BibitemOpen
  \bibfield  {author} {\bibinfo {author} {\bibfnamefont {X.}~\bibnamefont {Mi}}
  \emph {et~al.},\ }\href
  {https://doi.org/https://doi.org/10.1038/s41586-021-04257-w} {\bibfield
  {journal} {\bibinfo  {journal} {Nature}\ }\textbf {\bibinfo {volume} {601}},\
  \bibinfo {pages} {531} (\bibinfo {year} {2022})}\BibitemShut {NoStop}%
\bibitem [{\citenamefont {Stasiuk}\ and\ \citenamefont
  {Cappellaro}(2023)}]{prethermal_exp1}%
  \BibitemOpen
  \bibfield  {author} {\bibinfo {author} {\bibfnamefont {A.}~\bibnamefont
  {Stasiuk}}\ and\ \bibinfo {author} {\bibfnamefont {P.}~\bibnamefont
  {Cappellaro}},\ }\href {https://doi.org/10.1103/PhysRevX.13.041016}
  {\bibfield  {journal} {\bibinfo  {journal} {Phys. Rev. X}\ }\textbf {\bibinfo
  {volume} {13}},\ \bibinfo {pages} {041016} (\bibinfo {year}
  {2023})}\BibitemShut {NoStop}%
\bibitem [{\citenamefont {Kyprianidis}\ \emph {et~al.}(2021)\citenamefont
  {Kyprianidis}, \citenamefont {Machado}, \citenamefont {Morong}, \citenamefont
  {Becker}, \citenamefont {Collins}, \citenamefont {Else}, \citenamefont
  {Feng}, \citenamefont {Hess}, \citenamefont {Nayak}, \citenamefont {Pagano},
  \citenamefont {Yao},\ and\ \citenamefont {Monroe}}]{prethermal_exp2}%
  \BibitemOpen
  \bibfield  {author} {\bibinfo {author} {\bibfnamefont {A.}~\bibnamefont
  {Kyprianidis}}, \bibinfo {author} {\bibfnamefont {F.}~\bibnamefont
  {Machado}}, \bibinfo {author} {\bibfnamefont {W.}~\bibnamefont {Morong}},
  \bibinfo {author} {\bibfnamefont {P.}~\bibnamefont {Becker}}, \bibinfo
  {author} {\bibfnamefont {K.~S.}\ \bibnamefont {Collins}}, \bibinfo {author}
  {\bibfnamefont {D.~V.}\ \bibnamefont {Else}}, \bibinfo {author}
  {\bibfnamefont {L.}~\bibnamefont {Feng}}, \bibinfo {author} {\bibfnamefont
  {P.~W.}\ \bibnamefont {Hess}}, \bibinfo {author} {\bibfnamefont
  {C.}~\bibnamefont {Nayak}}, \bibinfo {author} {\bibfnamefont
  {G.}~\bibnamefont {Pagano}}, \bibinfo {author} {\bibfnamefont {N.~Y.}\
  \bibnamefont {Yao}},\ and\ \bibinfo {author} {\bibfnamefont {C.}~\bibnamefont
  {Monroe}},\ }\href {https://doi.org/10.1126/science.abg8102} {\bibfield
  {journal} {\bibinfo  {journal} {Science}\ }\textbf {\bibinfo {volume}
  {372}},\ \bibinfo {pages} {1192} (\bibinfo {year} {2021})}\BibitemShut
  {NoStop}%
\bibitem [{\citenamefont {Maskara}\ \emph {et~al.}(2021)\citenamefont
  {Maskara}, \citenamefont {Michailidis}, \citenamefont {Ho}, \citenamefont
  {Bluvstein}, \citenamefont {Choi}, \citenamefont {Lukin},\ and\ \citenamefont
  {Serbyn}}]{dtc_lukin}%
  \BibitemOpen
  \bibfield  {author} {\bibinfo {author} {\bibfnamefont {N.}~\bibnamefont
  {Maskara}}, \bibinfo {author} {\bibfnamefont {A.~A.}\ \bibnamefont
  {Michailidis}}, \bibinfo {author} {\bibfnamefont {W.~W.}\ \bibnamefont {Ho}},
  \bibinfo {author} {\bibfnamefont {D.}~\bibnamefont {Bluvstein}}, \bibinfo
  {author} {\bibfnamefont {S.}~\bibnamefont {Choi}}, \bibinfo {author}
  {\bibfnamefont {M.~D.}\ \bibnamefont {Lukin}},\ and\ \bibinfo {author}
  {\bibfnamefont {M.}~\bibnamefont {Serbyn}},\ }\href
  {https://doi.org/10.1103/PhysRevLett.127.090602} {\bibfield  {journal}
  {\bibinfo  {journal} {Phys. Rev. Lett.}\ }\textbf {\bibinfo {volume} {127}},\
  \bibinfo {pages} {090602} (\bibinfo {year} {2021})}\BibitemShut {NoStop}%
\bibitem [{\citenamefont {Choi}\ \emph {et~al.}(2019)\citenamefont {Choi},
  \citenamefont {Turner}, \citenamefont {Pichler}, \citenamefont {Ho},
  \citenamefont {Michailidis}, \citenamefont {Papi\ifmmode~\acute{c}\else
  \'{c}\fi{}}, \citenamefont {Serbyn}, \citenamefont {Lukin},\ and\
  \citenamefont {Abanin}}]{qmbs}%
  \BibitemOpen
  \bibfield  {author} {\bibinfo {author} {\bibfnamefont {S.}~\bibnamefont
  {Choi}}, \bibinfo {author} {\bibfnamefont {C.~J.}\ \bibnamefont {Turner}},
  \bibinfo {author} {\bibfnamefont {H.}~\bibnamefont {Pichler}}, \bibinfo
  {author} {\bibfnamefont {W.~W.}\ \bibnamefont {Ho}}, \bibinfo {author}
  {\bibfnamefont {A.~A.}\ \bibnamefont {Michailidis}}, \bibinfo {author}
  {\bibfnamefont {Z.}~\bibnamefont {Papi\ifmmode~\acute{c}\else \'{c}\fi{}}},
  \bibinfo {author} {\bibfnamefont {M.}~\bibnamefont {Serbyn}}, \bibinfo
  {author} {\bibfnamefont {M.~D.}\ \bibnamefont {Lukin}},\ and\ \bibinfo
  {author} {\bibfnamefont {D.~A.}\ \bibnamefont {Abanin}},\ }\href
  {https://doi.org/10.1103/PhysRevLett.122.220603} {\bibfield  {journal}
  {\bibinfo  {journal} {Phys. Rev. Lett.}\ }\textbf {\bibinfo {volume} {122}},\
  \bibinfo {pages} {220603} (\bibinfo {year} {2019})}\BibitemShut {NoStop}%
\bibitem [{\citenamefont {Moudgalya}\ \emph {et~al.}(2022)\citenamefont
  {Moudgalya}, \citenamefont {Bernevig},\ and\ \citenamefont
  {Regnault}}]{review_qmbs}%
  \BibitemOpen
  \bibfield  {author} {\bibinfo {author} {\bibfnamefont {S.}~\bibnamefont
  {Moudgalya}}, \bibinfo {author} {\bibfnamefont {B.~A.}\ \bibnamefont
  {Bernevig}},\ and\ \bibinfo {author} {\bibfnamefont {N.}~\bibnamefont
  {Regnault}},\ }\href {https://doi.org/10.1088/1361-6633/ac73a0} {\bibfield
  {journal} {\bibinfo  {journal} {Rep. Prog. Phys.}\ }\textbf {\bibinfo
  {volume} {85}},\ \bibinfo {pages} {086501} (\bibinfo {year}
  {2022})}\BibitemShut {NoStop}%
\bibitem [{\citenamefont {Wang}\ \emph {et~al.}(2024)\citenamefont {Wang},
  \citenamefont {Oshikawa}, \citenamefont {Kormos},\ and\ \citenamefont
  {Wu}}]{wangxiaoprb}%
  \BibitemOpen
  \bibfield  {author} {\bibinfo {author} {\bibfnamefont {X.}~\bibnamefont
  {Wang}}, \bibinfo {author} {\bibfnamefont {M.}~\bibnamefont {Oshikawa}},
  \bibinfo {author} {\bibfnamefont {M.}~\bibnamefont {Kormos}},\ and\ \bibinfo
  {author} {\bibfnamefont {J.}~\bibnamefont {Wu}},\ }\href
  {https://doi.org/10.1103/PhysRevB.110.195101} {\bibfield  {journal} {\bibinfo
   {journal} {Phys. Rev. B}\ }\textbf {\bibinfo {volume} {110}},\ \bibinfo
  {pages} {195101} (\bibinfo {year} {2024})}\BibitemShut {NoStop}%
\bibitem [{SM()}]{SM}%
  \BibitemOpen
  \href@noop {} {}\bibinfo {howpublished} {See Supplemental
  Material}\BibitemShut {NoStop}%
\bibitem [{\citenamefont {Bethe}(1931)}]{Bethe}%
  \BibitemOpen
  \bibfield  {author} {\bibinfo {author} {\bibfnamefont {H.}~\bibnamefont
  {Bethe}},\ }\href {https://doi.org/https://doi.org/10.1007/BF01341708}
  {\bibfield  {journal} {\bibinfo  {journal} {Zeitschrift für Physik}\
  }\textbf {\bibinfo {volume} {71}},\ \bibinfo {pages} {205} (\bibinfo {year}
  {1931})}\BibitemShut {NoStop}%
\bibitem [{\citenamefont {Wu}\ \emph {et~al.}(2008)\citenamefont {Wu},
  \citenamefont {Zhao}, \citenamefont {Chen},\ and\ \citenamefont
  {Zhang}}]{jianda_qgp}%
  \BibitemOpen
  \bibfield  {author} {\bibinfo {author} {\bibfnamefont {J.-d.}\ \bibnamefont
  {Wu}}, \bibinfo {author} {\bibfnamefont {M.-s.}\ \bibnamefont {Zhao}},
  \bibinfo {author} {\bibfnamefont {J.-l.}\ \bibnamefont {Chen}},\ and\
  \bibinfo {author} {\bibfnamefont {Y.-d.}\ \bibnamefont {Zhang}},\ }\href
  {https://doi.org/10.1103/PhysRevA.77.062114} {\bibfield  {journal} {\bibinfo
  {journal} {Phys. Rev. A}\ }\textbf {\bibinfo {volume} {77}},\ \bibinfo
  {pages} {062114} (\bibinfo {year} {2008})}\BibitemShut {NoStop}%
\bibitem [{\citenamefont {Xu}\ \emph {et~al.}(2018)\citenamefont {Xu},
  \citenamefont {Wu},\ and\ \citenamefont {Wu}}]{geodesic_curvature}%
  \BibitemOpen
  \bibfield  {author} {\bibinfo {author} {\bibfnamefont {C.}~\bibnamefont
  {Xu}}, \bibinfo {author} {\bibfnamefont {J.}~\bibnamefont {Wu}},\ and\
  \bibinfo {author} {\bibfnamefont {C.}~\bibnamefont {Wu}},\ }\href
  {https://doi.org/10.1103/PhysRevA.97.032124} {\bibfield  {journal} {\bibinfo
  {journal} {Phys. Rev. A}\ }\textbf {\bibinfo {volume} {97}},\ \bibinfo
  {pages} {032124} (\bibinfo {year} {2018})}\BibitemShut {NoStop}%
\bibitem [{\citenamefont {Takayoshi}\ \emph {et~al.}(2021)\citenamefont
  {Takayoshi}, \citenamefont {Wu},\ and\ \citenamefont {Oka}}]{jianda_scipost}%
  \BibitemOpen
  \bibfield  {author} {\bibinfo {author} {\bibfnamefont {S.}~\bibnamefont
  {Takayoshi}}, \bibinfo {author} {\bibfnamefont {J.}~\bibnamefont {Wu}},\ and\
  \bibinfo {author} {\bibfnamefont {T.}~\bibnamefont {Oka}},\ }\href
  {https://doi.org/10.21468/SciPostPhys.11.4.075} {\bibfield  {journal}
  {\bibinfo  {journal} {SciPost Phys.}\ }\textbf {\bibinfo {volume} {11}},\
  \bibinfo {pages} {075} (\bibinfo {year} {2021})}\BibitemShut {NoStop}%
\bibitem [{\citenamefont {Anderson}(1967)}]{anderson}%
  \BibitemOpen
  \bibfield  {author} {\bibinfo {author} {\bibfnamefont {P.~W.}\ \bibnamefont
  {Anderson}},\ }\href {https://doi.org/10.1103/PhysRevLett.18.1049} {\bibfield
   {journal} {\bibinfo  {journal} {Phys. Rev. Lett.}\ }\textbf {\bibinfo
  {volume} {18}},\ \bibinfo {pages} {1049} (\bibinfo {year}
  {1967})}\BibitemShut {NoStop}%
\bibitem [{\citenamefont {Heyl}(2015)}]{LE}%
  \BibitemOpen
  \bibfield  {author} {\bibinfo {author} {\bibfnamefont {M.}~\bibnamefont
  {Heyl}},\ }\href {https://doi.org/10.1103/PhysRevLett.115.140602} {\bibfield
  {journal} {\bibinfo  {journal} {Phys. Rev. Lett.}\ }\textbf {\bibinfo
  {volume} {115}},\ \bibinfo {pages} {140602} (\bibinfo {year}
  {2015})}\BibitemShut {NoStop}%
\bibitem [{\citenamefont {Wang}\ \emph {et~al.}(2025)\citenamefont {Wang},
  \citenamefont {He},\ and\ \citenamefont {Wu}}]{spectrum_entropy}%
  \BibitemOpen
  \bibfield  {author} {\bibinfo {author} {\bibfnamefont {X.}~\bibnamefont
  {Wang}}, \bibinfo {author} {\bibfnamefont {X.}~\bibnamefont {He}},\ and\
  \bibinfo {author} {\bibfnamefont {J.}~\bibnamefont {Wu}},\ }\href
  {https://arxiv.org/abs/2503.18396} {\bibfield  {journal} {\bibinfo  {journal}
  {arXiv:2503.18396}\ } (\bibinfo {year} {2025})}\BibitemShut {NoStop}%
\bibitem [{\citenamefont {Sun}\ \emph {et~al.}(2021)\citenamefont {Sun},
  \citenamefont {Yang}, \citenamefont {Wang}, \citenamefont {Zhou},
  \citenamefont {Su}, \citenamefont {Dai}, \citenamefont {Yuan},\ and\
  \citenamefont {Pan}}]{cold_atom}%
  \BibitemOpen
  \bibfield  {author} {\bibinfo {author} {\bibfnamefont {H.}~\bibnamefont
  {Sun}}, \bibinfo {author} {\bibfnamefont {B.}~\bibnamefont {Yang}}, \bibinfo
  {author} {\bibfnamefont {H.-Y.}\ \bibnamefont {Wang}}, \bibinfo {author}
  {\bibfnamefont {Z.-Y.}\ \bibnamefont {Zhou}}, \bibinfo {author}
  {\bibfnamefont {G.-X.}\ \bibnamefont {Su}}, \bibinfo {author} {\bibfnamefont
  {H.-N.}\ \bibnamefont {Dai}}, \bibinfo {author} {\bibfnamefont {Z.-S.}\
  \bibnamefont {Yuan}},\ and\ \bibinfo {author} {\bibfnamefont {J.-W.}\
  \bibnamefont {Pan}},\ }\href
  {https://doi.org/https://doi.org/10.1038/s41567-021-01277-1} {\bibfield
  {journal} {\bibinfo  {journal} {Nature Physics}\ }\textbf {\bibinfo {volume}
  {17}},\ \bibinfo {pages} {990} (\bibinfo {year} {2021})}\BibitemShut
  {NoStop}%
\bibitem [{\citenamefont {Klauser}\ \emph {et~al.}(2012)\citenamefont
  {Klauser}, \citenamefont {Mossel},\ and\ \citenamefont {Caux}}]{formfactor}%
  \BibitemOpen
  \bibfield  {author} {\bibinfo {author} {\bibfnamefont {A.}~\bibnamefont
  {Klauser}}, \bibinfo {author} {\bibfnamefont {J.}~\bibnamefont {Mossel}},\
  and\ \bibinfo {author} {\bibfnamefont {J.-S.}\ \bibnamefont {Caux}},\ }\href
  {https://doi.org/10.1088/1742-5468/2012/03/P03012} {\bibfield  {journal}
  {\bibinfo  {journal} {Journal of Statistical Mechanics: Theory and
  Experiment}\ }\textbf {\bibinfo {volume} {2012}},\ \bibinfo {pages} {P03012}
  (\bibinfo {year} {2012})}\BibitemShut {NoStop}%
\bibitem [{\citenamefont {Arecchi}\ \emph {et~al.}(1972)\citenamefont
  {Arecchi}, \citenamefont {Courtens}, \citenamefont {Gilmore},\ and\
  \citenamefont {Thomas}}]{disentangle}%
  \BibitemOpen
  \bibfield  {author} {\bibinfo {author} {\bibfnamefont {F.~T.}\ \bibnamefont
  {Arecchi}}, \bibinfo {author} {\bibfnamefont {E.}~\bibnamefont {Courtens}},
  \bibinfo {author} {\bibfnamefont {R.}~\bibnamefont {Gilmore}},\ and\ \bibinfo
  {author} {\bibfnamefont {H.}~\bibnamefont {Thomas}},\ }\href
  {https://doi.org/10.1103/PhysRevA.6.2211} {\bibfield  {journal} {\bibinfo
  {journal} {Phys. Rev. A}\ }\textbf {\bibinfo {volume} {6}},\ \bibinfo {pages}
  {2211} (\bibinfo {year} {1972})}\BibitemShut {NoStop}%
\end{thebibliography}%

\newpage
\appendix
\setcounter{figure}{0}
\setcounter{equation}{0}
\renewcommand{\thefigure}{S\arabic{figure}}
\renewcommand{\theequation}{S\arabic{equation}}
\renewcommand{\theHfigure}{S.\arabic{figure}} 
\onecolumngrid
\section{\large{Supplemental Material---Discrete time crystal and perfect many-body tunneling in a periodically driven Heisenberg spin chain}}

\section*{A.~Periodic oscillation of magnetization and spin correlation}
This section calculates the magnetization $M_z$ and equal-time spin correlations for general initial state.
For $M_z$, we have
\begin{equation}
\braket{M_{z}}=\frac{1}{N}\bra{\psi(t)}\sum_{j=1}^{N}S_{j}^{z}\ket{\psi(t)}=\frac{1}{N}\sum_{j=1}^{N}\bra{\psi(0)}\text{exp}(iH_{\text{eff}}t)U_{D}(t)S_{j}^{z}U_{D}(t)^{\dagger}\text{exp}(-iH_{\text{eff}}t)\ket{\psi(0)}
\label{eq:magnetization}
\end{equation}
Because $[U_D(t), S_{j}^{z}]=0\ \text{for}\ \forall j$, $U_{D}(t)S_{j}^{z}U_{D}(t)^{\dagger}=S_{j}^{z}$. 
Since $[H_{\text{xxx}},\sum_jS_{j}^{z}]=0$ and 
\begin{equation}
\begin{aligned}
    & \text{exp}(i\left\{\sum_{j=1}^{N}h_rS_{j}^{x}+(\omega-h)S_{j}^{z}\right\}t)S_{j}^{z}\text{exp}(-i\left\{\sum_{j=1}^{N}h_rS_{j}^{x}+(\omega-h)S_{j}^{z}\right\}t) \\
    &= \left[\text{cos}^2(\vartheta)\text{cos}(\tilde{h}t)+\text{sin}^{2}(\vartheta)\right]S_{j}^{z}+\text{sin}(\vartheta)\text{cos}(\vartheta)\left[1-\text{cos}(\tilde{h}t)\right]S_{j}^{x}+\text{cos}(\vartheta)\text{sin}(\tilde{h}t)S_{j}^{y}\\
    &\equiv\vec{f}(t)\cdot\vec{S}_{j}
\end{aligned}
\end{equation}
where $\mathrm{tan}(\vartheta)\equiv\frac{\omega-h}{h_r}$ and $\tilde{h}=\sqrt{h_r^2+(\omega-h)^2}$, 
we have $M_z=1/N\vec{f}(t)\cdot\bra{\psi(0)}\sum_{j}\vec{S}_{j}\ket{\psi(0)}$. 
\par
For special case where initial state is the ground state of $H(0)$, the expression can be further reduced. 
Denote $\frac{1}{N}\sum_{j=1}^{N}\bra{GS(0)}S_{j}^{\mu}\ket{GS(0)}$ as $\braket{M_{\mu}}_{0}$, 
where $\mu=x,y,z$. Since the magnetic field of $H(0)$ lies in the $x\text{-}z$ plane, $\braket{M_y}_0=0$. We have
\begin{equation}
M_{z}=\left[\text{cos}^2(\vartheta)\text{cos}(\tilde{h}t)+\text{sin}^{2}(\vartheta)\right]\braket{M_{z}}_{0}+\text{sin}(\vartheta)\text{cos}(\vartheta)\left[1-\text{cos}(\tilde{h}t)\right]\braket{M_{x}}_{0}
\end{equation}

\begin{figure}
    \centering
    \includegraphics[width=0.98\linewidth]{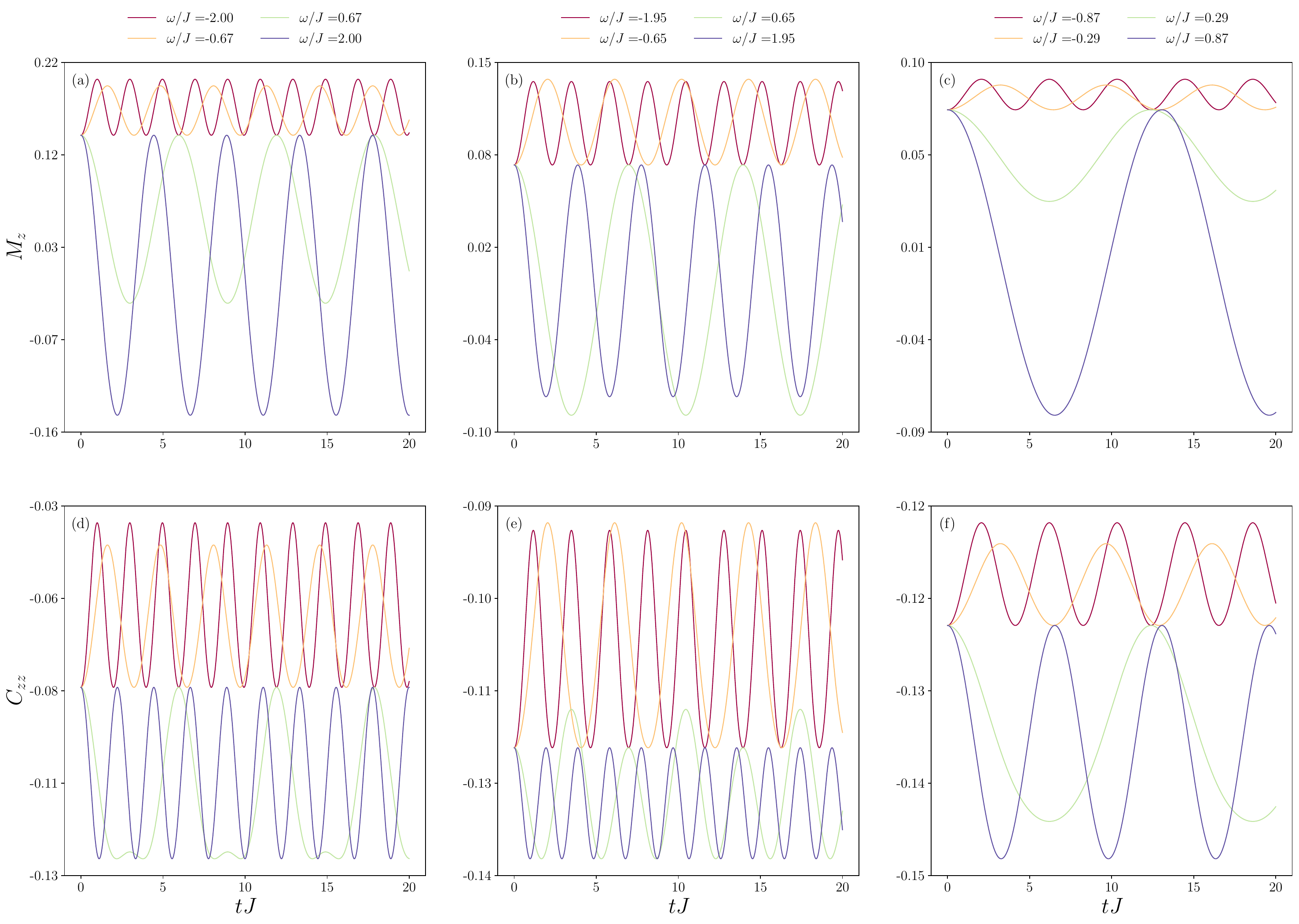}
    \caption{ (a), (b), (c), $M_z$ $vs.$ $tJ$; (d), (e), (f), $C_{zz}$ $vs.$ $tJ$ with $N=64$. In each sub-figure, the blue line represents the case where perfect MBT occurs. }
    \label{fig:more_observables}
\end{figure}

For the equal-time spin correlation of neighboring spins, it follows
\begin{equation}
    C_{zz} = \frac{1}{N}\sum_{j=1}^{N}\bra{\psi(t)}S_{j}^{z}S_{j+1}^{z}\ket{\psi(t)}
\end{equation}
Using the same procedure, we reach
\begin{equation}
    C_{zz} = \frac{1}{N}\sum_{j=1}^{N}\bra{\psi(0)}\text{exp}(i\left\{\sum_{j=1}^{N}h_rS_{j}^{x}+(\omega-h)S_{j}^{z}\right\}t)S_{j}^{z}S_{j+1}^{z}\text{exp}(-i\left\{\sum_{j=1}^{N}h_rS_{j}^{x}+(\omega-h)S_{j}^{z}\right\}t)\ket{\psi(0)}
\end{equation}
Denote $\cos\alpha S_{j}^{x}-\sin\alpha S_j^{z}$ as $\tilde{S}_{j}^{z}$, $\sin\alpha S_{j}^{x}+\cos\alpha S_j^{z}$ as $\tilde{S}_{j}^{x}$ and $S_j^{y}$ as $\tilde{S}_{j}^{y}$ where $\cos\alpha \equiv h_r/\sqrt{h_r^2+h^2},\sin\alpha\equiv h/\sqrt{h_r^2+h^2}$, we have 
\begin{equation}
    \vec{f}(t)\cdot\vec{S}_{j}=\mathsf{A}(t)\tilde{S}_j^{x}+\mathsf{B}(t)\tilde{S
    }_{j}^{y}+\mathsf{C}(t)\tilde{S}_{j}^{z}
\end{equation}
where
\begin{equation}
    \begin{aligned}
\mathsf{A}(t)&\equiv\left[\cos(\tilde{h}t)\cos(\vartheta)\cos(\alpha+\vartheta)+\sin(\vartheta)\sin(\alpha+\vartheta)\right]\\
\mathsf{B}(t)&\equiv\text{cos}(\vartheta)\text{sin}(\tilde{h}t)\\
\mathsf{C}(t)&\equiv\left[\cos(\tilde{h}t)\cos(\vartheta)\sin(\alpha+\vartheta)-\cos(\alpha+\vartheta)\sin(\vartheta)\right]
    \end{aligned}
\end{equation}
and $\mathsf{A}^2(t)+\mathsf{B}^2(t)+\mathsf{C}^2(t)=1$.
$C_{zz}$ can be written as 
\begin{equation}
    C_{zz}=\frac{1}{N}\sum_{j=1}^{N}\bra{\psi(0)}\left[\vec{f}(t)\cdot\vec{S}_{j}\right]\left[\vec{f}(t)\cdot\vec{S}_{j+1}\right]\ket{\psi(0)}.
\end{equation}
This can be also reduced for the initial state being $\ket{GS(0)}$ case.  
Denote $\frac{1}{N}\bra{GS(0)}\sum_{j=1}^{N}\tilde{S}_{j}^{\mu}\tilde{S}_{j+1}^{\nu}\ket{GS(0)}$ as $\braket{C_{\mu\nu}}_{0}$. $\braket{C_{xy}}_0,\braket{C_{xz}}_0$ and $\braket{C_{yz}}_0$ are 0.
Due to the $U(1)$ symmetry, $\braket{C_{xx}}_{0}=\braket{C_{yy}}_{0}$. We have
\begin{equation}
    C_{zz}=\braket{C_{xx}}_0+\mathsf{C}^2(t)\left(\braket{C_{zz}}_0-\braket{C_{xx}}_0\right)
\end{equation}
where $\braket{C_{zz}}_0$ and $\braket{C_{xx}}_0$ are determined using the form factor formula \cite{formfactor}. 
In general the period of $C_{zz}$ is $T_{\text{re}}$. There are two exceptions: $\vartheta=0$ and $\alpha+\vartheta=\pi/2$ with period $T_{\text{re}}/2$. They correspond to 
$h=\omega$ and Eq.~\eqref{eq:square} respectively. The $C_{zz}$ also exhibits the
discrete TTSB [Fig.~\ref{fig:more_observables}]. 
\section*{B.~Calculation of Loschmidt echo and Spectrum entropy}
In this section, we introduce the method of calculating Loschmidt echo $\mathcal{L}(t)$ and spectrum entropy. From the discussion above we know that $\mathcal{L}(t)=\left|\bra{GS(0)}e^{-iH_{\text{eff}}t}\ket{GS(0)}\right|^2$. Inserting the complete basis of $H_{\text{eff}}$, we have:
\begin{equation}
\begin{aligned}
\mathcal{L}(t)&=\left|\sum_{n}\braket{GS(0)|n}\bra{n}e^{-iH_{eff}t}\ket{GS(0)}\right|^2\\
&=\sum_{n,m}\mathcal{O}_n\mathcal{O}_m e^{-i(E_{n}-E_{m})t}
\label{eq:sm_le}
\end{aligned}
\end{equation}
where $\mathcal{O}_n\equiv|\braket{GS(0)|n}|^2$. 
We then introduce a unitary transformation $U_\alpha=\text{exp}\left(-i\alpha\sum_{j=1}^{N}S_{j}^{y}\right)$ which satisfies
\begin{equation}
    U_\alpha\sum_{j=1}^{N}\left(h_rS_{j}^{x}-hS_{j}^{z}\right)U_\alpha^{\dagger}=\sqrt{h_r^2+h^2}\sum_{j=1}^{N}\left(\frac{h_r}{\sqrt{h_r^2+(\omega-h)^2}}S_{j}^{x}+\frac{\omega-h}{\sqrt{h_r^2+(\omega-h)^2}}S_{j}^{z}\right)
    \label{eq:ur}
\end{equation}
The angle $\alpha$ is determined as $\text{tan}\alpha=-h_r\omega/(h^2+h_r^2-h\omega)$.
Eq.~\eqref{eq:ur} shows that the eigenstates of $H_{\text{eff}}$ can be generated by applying $U_\alpha$ on the eigenstates of $H(0)$. Then $\mathcal{O}_{n}=\left|\bra{GS(0)}U_\alpha\ket{\tilde{n}}\right|^2$, where $\ket{\tilde{n}}=U_\alpha^{\dagger}\ket{n}$ is eigenstate of $H(0)$. Let $2\mathsf{j}+1$ be the dimension of the irreducible representation to which $\ket{GS(0)}$ belongs. Denoting the SU(2) index of $\ket{\tilde{n}}$ as $\mathsf{n}$, we have
\begin{equation}
    \mathcal{O}_{n}=|\bra{\mathsf{j,n}}\text{exp}\left\{-\frac{\alpha}{2}\sum_{j=1}^{N}(S_{j}^{+}-S_{j}^{-})\right\}\ket{\mathsf{j,j}}|^2
    \label{eq:overlap}
\end{equation}

From now on, we denote $\mathcal{O}_{n}$ as $\mathcal{O}_{\alpha,\mathsf{n}}$ and $\mathcal{L}(t)$ as $\mathcal{L}_{\alpha}(t)$.
For convenience, we write $\sum_{j=1}^{N}S_{j}^{+}$ as $J_{+}$, $\sum_{j=1}^{N}S_{j}^{-}$ as $J_-$ and $\sum_{j=1}^{N}S_{j}^{z}$ as $J_{z}$. We then have 
\begin{equation}
    \mathcal{O}_{\alpha,\mathsf{n}}=|\bra{\mathsf{j,n}}\exp\left\{-\frac{\alpha}{2} (J_{+}-J_{-})\right\}\ket{\mathsf{j,j}}|^2
\end{equation}
By using the disentangling theorem \cite{disentangle}, we have
\begin{equation}
    \exp\left\{-\frac{\alpha}{2}(J_{+}-J_{-})\right\}=\exp(-\tau J_{-})\exp(\ln(1+|\tau|^2)J_{z})\exp(\tau J_{+})
\end{equation}
where $\tau\equiv-\tan\frac{\alpha}{2}$. Since $J_{+}\ket{\mathsf{j,j}}=0$, we have 
\begin{equation}
    \mathcal{O}_{\alpha,\mathsf{n}}=\frac{\tau^{2(\mathsf{j-n})}}{(1+\tau^2)^{2\mathsf{j}}}\prod_{\mathsf{p}=\mathsf{n}+1}^{\mathsf{j}}\frac{\mathsf{j}(\mathsf{j}+1)-\mathsf{p}(\mathsf{p}-1)}{(\mathsf{p-n})^2}
\end{equation}
$\mathcal{O}_{\alpha,\mathsf{n+1}}/\mathcal{O}_{\alpha,\mathsf{n}}=(\mathsf{j-n})/(\mathsf{j+n+}1)/\tau^2$. Let $\mathsf{q}=\mathsf{n+j}$, $\mathcal{O}_{\alpha,\mathsf{n+1}}/\mathcal{O}_{\alpha,\mathsf{n}}=(\mathsf{2j-q})/(\mathsf{q+}1)/\tau^2$, 
which is identical to $C_{2\mathsf{j}}^{\mathsf{q+1}}/C_{2\mathsf{j}}^{\mathsf{q}}/\tau^2$. Since $\mathcal{O}_{\alpha,\mathsf{-j}}=\tau^{4\mathsf{j}}/(1+\tau^2)^{2\mathsf{j}}$=$\tau^{4\mathsf{j}}C_{2\mathsf{j}}^{\mathsf{0}}/(1+\tau^2)^{2\mathsf{j}}$. We draw the conclusion that $\mathcal{O}_{\alpha,\mathsf{n}}=\tau^{2(\mathsf{j-n})}C_{2\mathsf{j}}^{\mathsf{n+j}}/(1+\tau^2)^{2\mathsf{j}}$. 
Because $E_n=\braket{n|H_{\text{eff}}|n}=\braket{\tilde{n}|U_\alpha^{\dagger}H_{\text{eff}}U_\alpha|\tilde{n}}$. We have
\begin{equation}
\begin{aligned}
    E_n-E_m&=\braket{\tilde{n}|U_\alpha^{\dagger}H_{\text{eff}}U_\alpha|\tilde{n}}-\braket{\tilde{m}|U_\alpha^{\dagger}H_{\text{eff}}U_\alpha|\tilde{m}}\\
    &=\tilde{h}(\mathsf{n-m})
    \end{aligned}
\end{equation}
Plug $\mathcal{O}_{\alpha, \mathsf{n}}$ into Eq.~\eqref{eq:sm_le},
\begin{equation}
\begin{aligned}
    \mathcal{L}(t)&=\sum_{\mathsf{n=-j}}^{\mathsf{j}}\sum_{\mathsf{m=-j}}^{\mathsf{j}}\mathcal{O}_{\alpha,\mathsf{n}}\mathcal{O}_{\alpha,\mathsf{m}}\text{cos}(\tilde{h}(\mathsf{n-m})t)\\
    &=\frac{\tau^{8\mathsf{j}}}{(1+\tau^2)^{4\mathsf{j}}}\sum_{\mathsf{p}=0}^{2\mathsf{j}}\sum_{\mathsf{q}=0}^{2\mathsf{j}}C_{\mathsf{2j}}^{\mathsf{p}}C_{\mathsf{2j}}^{\mathsf{q}}e^{i\tilde{h}t\mathsf{p}}/\tau^{2\mathsf{p}}e^{-i\tilde{h}t\mathsf{q}}/\tau^{2\mathsf{q}}\\
    &=\frac{\tau^{8\mathsf{j}}}{(1+\tau^2)^{4\mathsf{j}}}\left(1+e^{i\tilde{h}t}/\tau^2\right)^{2\mathsf{j}}\left(1+e^{-i\tilde{h}t}/\tau^2\right)^{2\mathsf{j}}\\
    &=\frac{\left(\tau^4+2\tau^2\cos\tilde{h}t+1\right)^{2\mathsf{j}}}{(1+\tau^2)^{4\mathsf{j}}}
\end{aligned}
\end{equation}
Following the definition of $S_\alpha$, we have
\begin{equation}
    S_{\alpha}=-\sum_{\mathsf{n=-j}}^{\mathsf{j}}\sum_{\mathsf{m=-j}}^{\mathsf{j}}\mathcal{O}_{\alpha,\mathsf{n}}\mathcal{O}_{\alpha,\mathsf{m}}\text{ln}(\mathcal{O}_{\alpha,\mathsf{n}}\mathcal{O}_{\alpha,\mathsf{m}})=-2\sum_{\mathsf{n=-j}}^{\mathsf{j}}\mathcal{O}_{\alpha,\mathsf{n}}\ln \mathcal{O}_{\alpha,\mathsf{n}}.
\end{equation}

\section{C.~ Introduction on Bethe ansatz}
This section gives an introduction on coordinate Bethe ansatz \cite{Bethe}. The XXX Hamiltonian is given by
\begin{equation}
    H_{\mathrm{xxx}}=J\sum_{j=1}^{N}\left(S_{j}^{x}S_{j+1}^{x}+S_{j}^{y}S_{j+1}^{y}+S_{j}^{z}S_{j+1}^{z}-\frac{1}{4}\right)
\end{equation}
$H_{\mathrm{xxx}}$ is SU(2) invariant, which means that eigenstates of $H_{\mathrm{xxx}}$
can be classified by SU(2) irreducible representations.
Bethe proposed 
that we can use magnon excitations to describe different eigenstates. Since periodic boundary condition
is applied, each magnon has well defined crystal momentum. We define the rapidity $\theta_{j}$ of 
$j^{th}$ magnon whose crystal momentum is denoted by $p_{j}$,
\begin{equation}
    e^{ip_{j}}=\frac{\theta_j+\frac{i}{2}}{\theta_j-\frac{i}{2}}
    \label{eq:momentum}
\end{equation} 
If a state with $M$ magnons is an eigenstate of $H_{\mathrm{xxx}}$, the rapidity of each magnon can
be solved from the Bethe ansatz equations,
\begin{equation}
\left(\frac{\theta_j+\frac{i}{2}}{\theta_j-\frac{i}{2}}\right)^N=
\prod_{k\neq j}^{M}\frac{\theta_j-\theta_k+i}{\theta_j-\theta_k-i}
\label{eq:bae}
\end{equation}
The corresponding eigenenergy follows by
\begin{equation}
    E({\theta})=-J\sum_{j=1}^{M}\frac{2}{4\theta_j^{2}+1}
    \label{eq:energy}
\end{equation}
Apparently, if $\left\{ \theta\right\}$ is rapidity set of an eigenstate, 
so is set $\left\{ \theta, \infty \right\}$. Eq.~\eqref{eq:momentum}
and Eq.~\eqref{eq:energy} imply that crystal momentum of magnon with unbound rapidity is zero and 
two eigenstates $\ket{\left\{ \theta\right\}}$ and $\ket{\left\{ \theta, \infty \right\}}$ are 
degenerate. The relation between $\ket{\left\{ \theta\right\}}$ and $\ket{\left\{ \theta, \infty \right\}}$
is that $\ket{\left\{ \theta, \infty \right\}}\propto \sum_{j=1}^{N}S_{j}^{-}\ket{\left\{ \theta\right\}}$, which 
implies that each eigenstate containing only finite norm rapidities (rapidity can be complex with
finite norm) is highest weight state of a SU(2) irreducible
representation. And other states in this representation are generated by acting $\sum_{j=1}^{N}S_{j}^{-}$ on the 
highest weight state. 

\end{document}